\documentclass[fleqn,usenatbib]{mnras}

\usepackage{newtxtext,newtxmath}

\usepackage[T1]{fontenc}
\usepackage{ae,aecompl}

\usepackage{graphicx}	
\usepackage{amsmath}	
\usepackage{amssymb}	

\title[The VISTA Photometric System] {The VISTA $Z$$Y\!$$J$$H$$Ks$ Photometric System: Calibration from 2MASS}

\author[Gonz\'alez-Fern\'andez et al.]
{C. Gonz\'alez-Fern\'andez$^{1}$,
S. T. Hodgkin$^{1}$,
M. J. Irwin$^{1}$,
E. Gonz\'{a}lez-Solares$^{1}$,
\newauthor
S. E. Koposov$^{2,1}$,
J. R. Lewis$^{1}$,
J. P. Emerson$^{3}$,
P. C. Hewett$^{1}$,
\newauthor
A. K. Yolda\c{s}$^{1}$,
M. Riello$^{1}$
\\
$^1$Institute of Astronomy, Madingley Road, Cambridge CB3 0HA, UK\\
$^2$McWilliams Center for Cosmology, Department of Physics, Carnegie
Mellon University, 5000 Forbes Avenue, Pittsburgh, PA 15213, USA\\
$^3$Astronomy Unit, School of Physics \& Astronomy, Queen Mary
University of London, Mile End Road, London E1 4NS, UK
}

\date{Accepted XXX. Received YYY; in original form ZZZ}

\pubyear{2015}

\begin{document}
\label{firstpage}
\pagerange{\pageref{firstpage}--\pageref{lastpage}}
\maketitle

\begin{abstract}
In this paper we describe the routine photometric calibration of data taken with the VIRCAM instrument on the ESO VISTA telescope. The broadband ZYJHKs data are directly calibrated from 2MASS point sources visible in every VISTA image. We present the empirical transformations between the 2MASS and VISTA, and WFCAM and VISTA, photometric systems for regions of low reddening. We investigate the long-term performance of VISTA+VIRCAM. An investigation of the dependence of the photometric calibration on interstellar reddening leads to these conclusions: (1) For all broadband filters, a linear colour-dependent correction compensates the gross effects of reddening where $E(B-V)<5.0$. (2) For $Z$ and $Y$, there is a significantly larger scatter above E(B-V)=5.0, and insufficient measurements to adequately constrain the relation beyond this value. (3) The $JHK\!s$ filters can be corrected to a few percent up to E(B-V)=10.0. We analyse spatial systematics over month-long timescales, both inter- and intra-detector and show that these are present only at very low levels in VISTA. We monitor and remove residual detector-to-detector offsets. We compare the calibration of the main pipeline products: pawprints and tiles. We show how variable seeing and transparency affect  the final calibration accuracy of VISTA tiles, and discuss a technique, {\it grouting}, for mitigating these effects. Comparison between repeated reference fields is used to demonstrate that the VISTA photometry is precise to better than $\simeq2\%$ for the $Y$$J$$H$$Ks$ bands and $3\%$ for the $Z$ bands. Finally we present empirically determined offsets to transform VISTA magnitudes into a true Vega system.
\end{abstract}




\section{Introduction}

\begin{figure}
\begin{center}
\includegraphics[height=7cm]{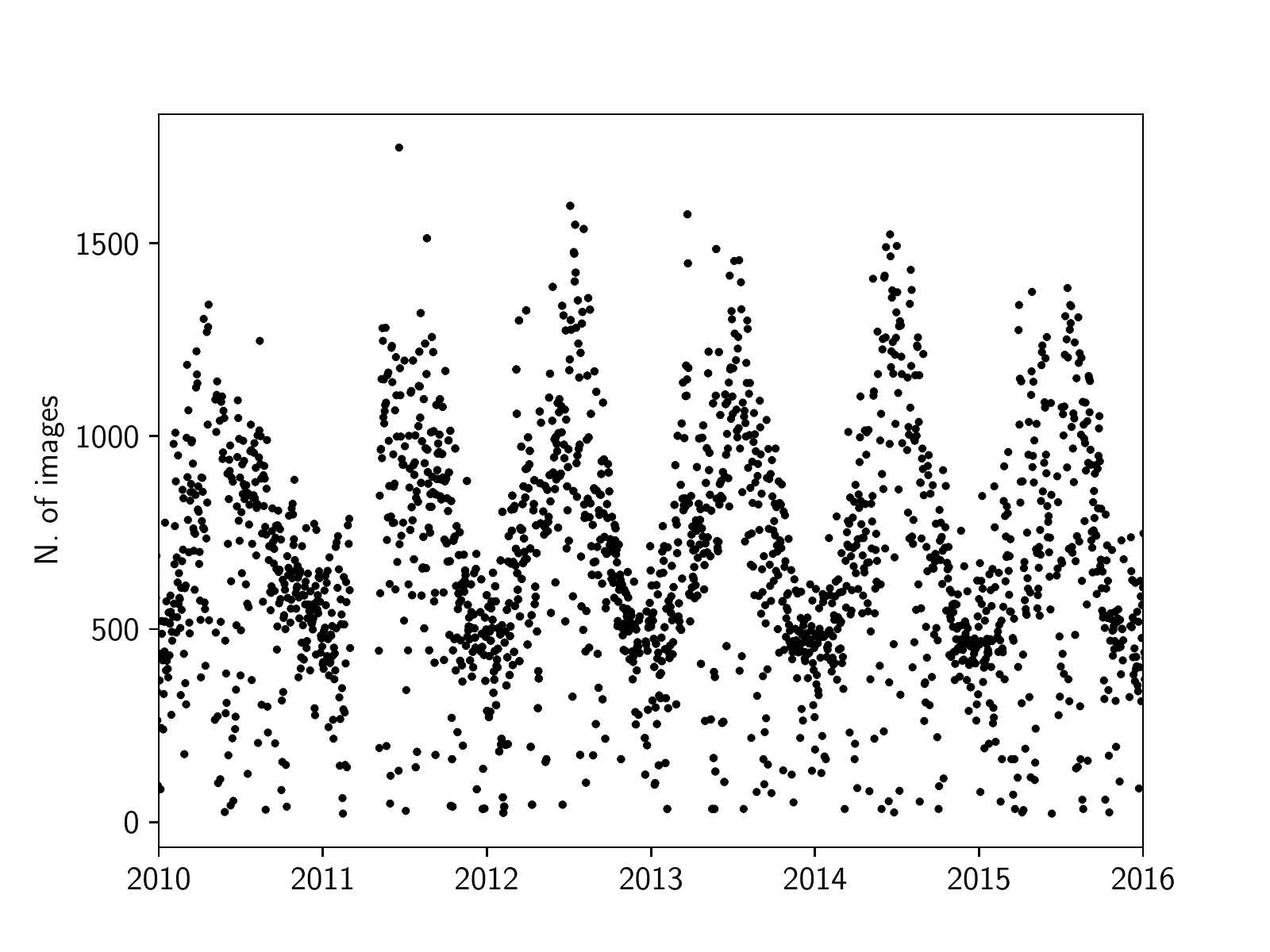}
\caption{Number of raw images per night taken with VISTA. The yearly modulation is mostly due to the visibility of the Galactic plane.}
\label{histn}
\end{center}
\end{figure}

The Visible and Infrared Survey Telescope for Astronomy (VISTA) is a 4-m telescope designed specifically for imaging survey work at visible and near-infrared wavelengths \citep{sut15}.  The VISTA infrared camera (VIRCAM) has a field of view of $1.65^\circ$ diameter with a pawprint that covers 0.59 degree$^2$ in the ZYJHKs passbands, using a 4$\times$4 array of 2k$\times$2k non-buttable Raytheon detectors with 0.34 arcsec pixels resulting in a 269MB FITS file per exposure.  The camera produces an average of 500GB of data a night with a high volume nights producing up to $\approx$2TB, in particular when the Galactic plane is accessible from Paranal. Fig. \ref{histn} shows the evolution of the number of raw sky images per night taken with VISTA over the years.

All VISTA data are processed at the Cambridge Astronomical Survey Unit (CASU) using software developed under the broad title of the VISTA Data Flow System (VDFS). VDFS has a broad remit and was setup to deliver:
\begin{enumerate}
\item a system for evaluating data quality run at the telescope in Paranal;
\item a system for monitoring the technical performance and producing quick 
look calibrated data at ESO Garching;
\item a system for delivering science-ready photometrically and 
astrometrically calibrated catalogues and images run in Cambridge;
\item a VISTA Science Archive designed and run by the Wide Field
Astronomy Unit (WFAU) in Edinburgh.
\end{enumerate}
The main scientific pipeline for VISTA will be discussed in Gonz\'alez-Fern\'andez et al. 2017, while the VISTA science Archive is described in \citet{cro12}. In this paper we focus specifically on the photometric calibration which is an evolution of strategies and software already developed for near-infrared data measured with the United Kingdom Infrared Telescope (UKIRT) Wide-Field Camera (WFCAM), and described in \citep[][hereafter H09]{hod09}.

Although as noted we are mainly concerned here with the photometric calibration several other aspects of the processing architecture impact this and these are also discussed as required. The overall pipeline processing architecture is designed to produce well-calibrated quality-assessed science products. To meet this aim, the end-to-end system must robustly remove instrument and night sky signatures; monitor data quality and system integrity; provide astrometric and photometric calibration; and generate photon noise-limited images and astronomical catalogues. 

The primary photometric calibrators are drawn from stars in the Two Micron all Sky Survey \citep[2MASS,][]{skr06}, present in large numbers in every VIRCAM science exposure (the median number is around 200 per pointing, spread across the 16 detectors). The 2MASS calibration has been shown to be uniform across the whole sky to better than 2\,per cent accuracy \citep{nik00}.

We have already demonstrated with WFCAM on the United Kingdom InfraRed Telescope \citep{cas07} that an accurate calibration may be derived using 2MASS (H09). VISTA, despite obvious design differences, represents a similar challenge: it is a near infrared large-format camera, with a similar sensitivity and spatial resolution to WFCAM, generating (comparably) large volumes of data on a nightly basis. All VISTA survey programmes to date have chosen to adopt a default pipeline photometric calibration\footnote{ESO also observe a number of reference fields every night to ensure a backup calibration strategy is available.}. This paper describes and discusses that method.

The paper is organised as follows: 

\begin{enumerate}
\item Introduction to the VISTA+VIRCAM system, observation strategies and overview of the reduction pipeline (Sections 1-3).
\item Calibration against 2MASS, derivation of colour equations and strategies to deal with extinction (Section 4).
\item Monitoring sensitivity, comparison with WFCAM photometric system and overall performance:
internal consistency, structure of the residuals, transforming from VISTA into true Vega magnitudes (Sections 5-7).
\item Conclusions.
\item Appendices.
\end{enumerate}

\section{VISTA and VIRCAM}

\begin{figure}
\begin{center}
\includegraphics[width=8.3cm]{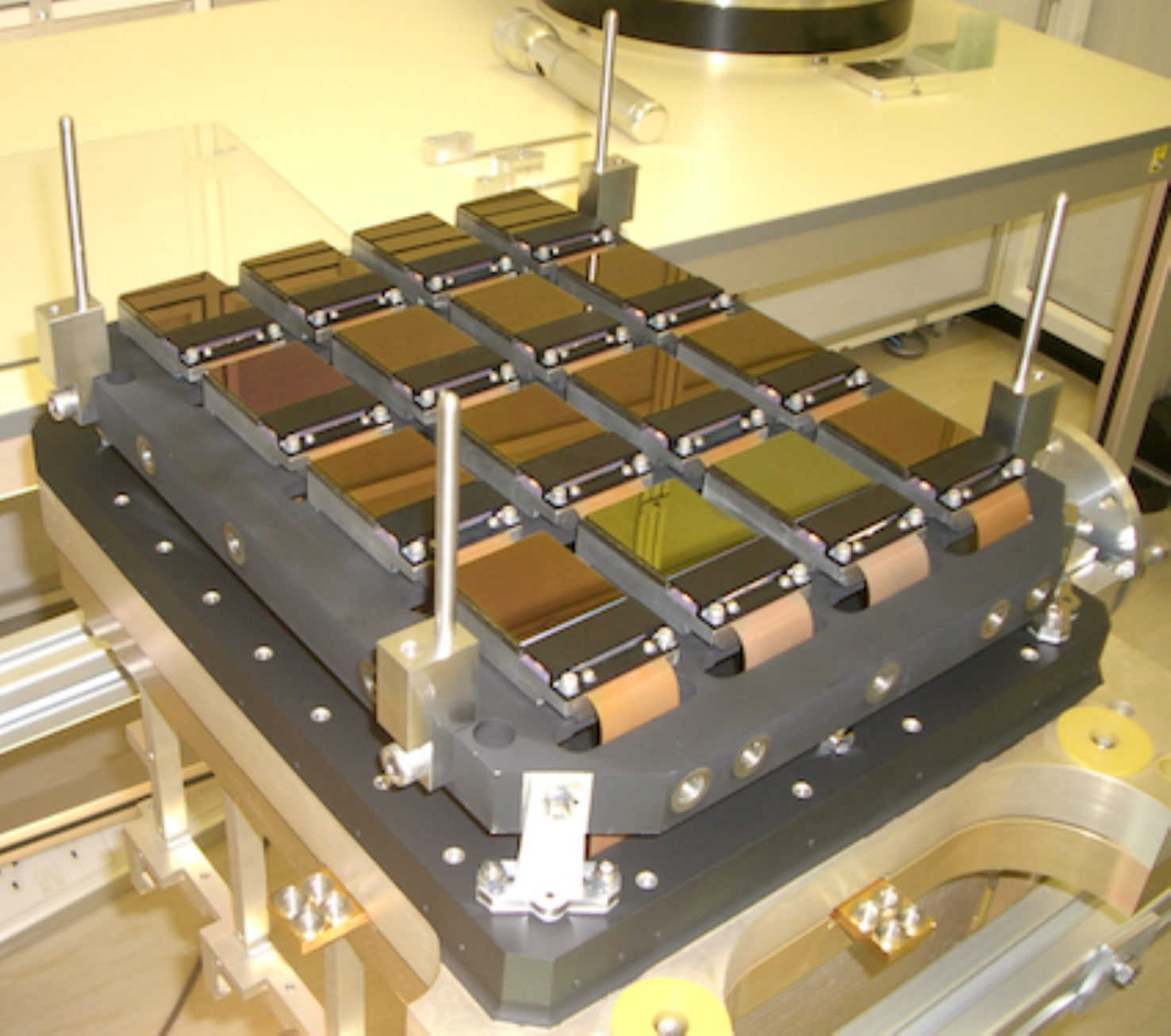}
\includegraphics[width=8.3cm]{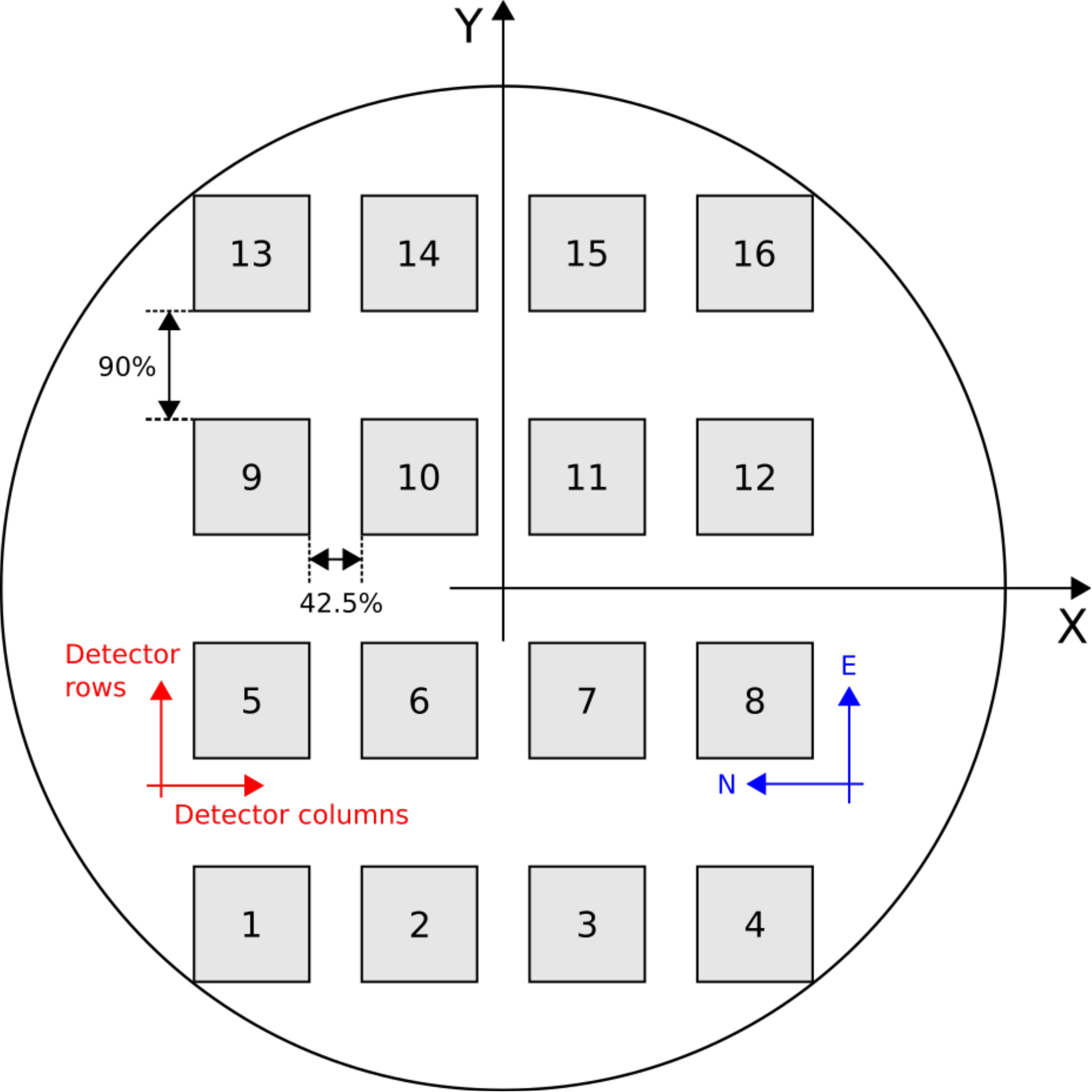}
\caption{{\bf Top:} The focal plane of VIRCAM showing the detector readout electronics at the bottom of each detector. {\bf Botttom:} Schematic view of the array unit. Each Raytheon VIRGO detector is 2048$\times$2048 pixels in size, approx. 11.6' in the sky. The (X,Y) convention followed in this paper is indicated, with X increasing across detector columns and Y across rows; in these coordinates, for the nominal rotator angle, East is up and North left. Readout amplifiers are adjacent to the bottom row of each detector. The Y axis points towards the centre of the filter wheel.}
\label{fplane}
\end{center}
\end{figure}

As can be seen in Fig. \ref{fplane}, the detector array in VISTA does not offer contiguous coverage, with gaps of $\sim42\%$ and $\sim90\%$ of a detector length between elements. To achieve contiguous sky coverage, VISTA usually observes following a 6 step pointing pattern that fills (tiles) about $1.5\,\mathrm{deg^2}$ (this process is outlined in Fig. \ref{tileproc}). At each pointing, two or more dither (jitter) exposures are taken, with a small spatial offset between them (a few arcseconds typically), to correct for small scale cosmetics. We refer to the combination of these single exposures, or pawprints, as a stacked pawprint image. The continuous mosaic resulting from combining 6 of these pawprint stacks we call a tile. For each one of these images, routine pipeline processing at CASU yields both the image itself and a photometric catalogue.

Each detector uses 16 contiguous channels that are read in parallel through separate amplifiers. One aspect of VIRCAM that is different to WFCAM is that significant (1--10\%) non-linearity over the full ADU range is present in all detectors and in all amplifier readouts. This is monitored on a regular monthly basis and corrections for the non-linearity applied at the front-end of all processing. A further complicating factor is the range of saturation levels present on each detector (24000ADU -- 37000ADU), which for tiled images forces a conservative choice of saturation level. A full account of these corrections along with the procedures involved in generating science-ready images and catalogues will be presented in a future paper (Gonz\'alez-Fern\'andez et al. 2017). 

\begin{figure}
\begin{center}
 \includegraphics[width=9cm]{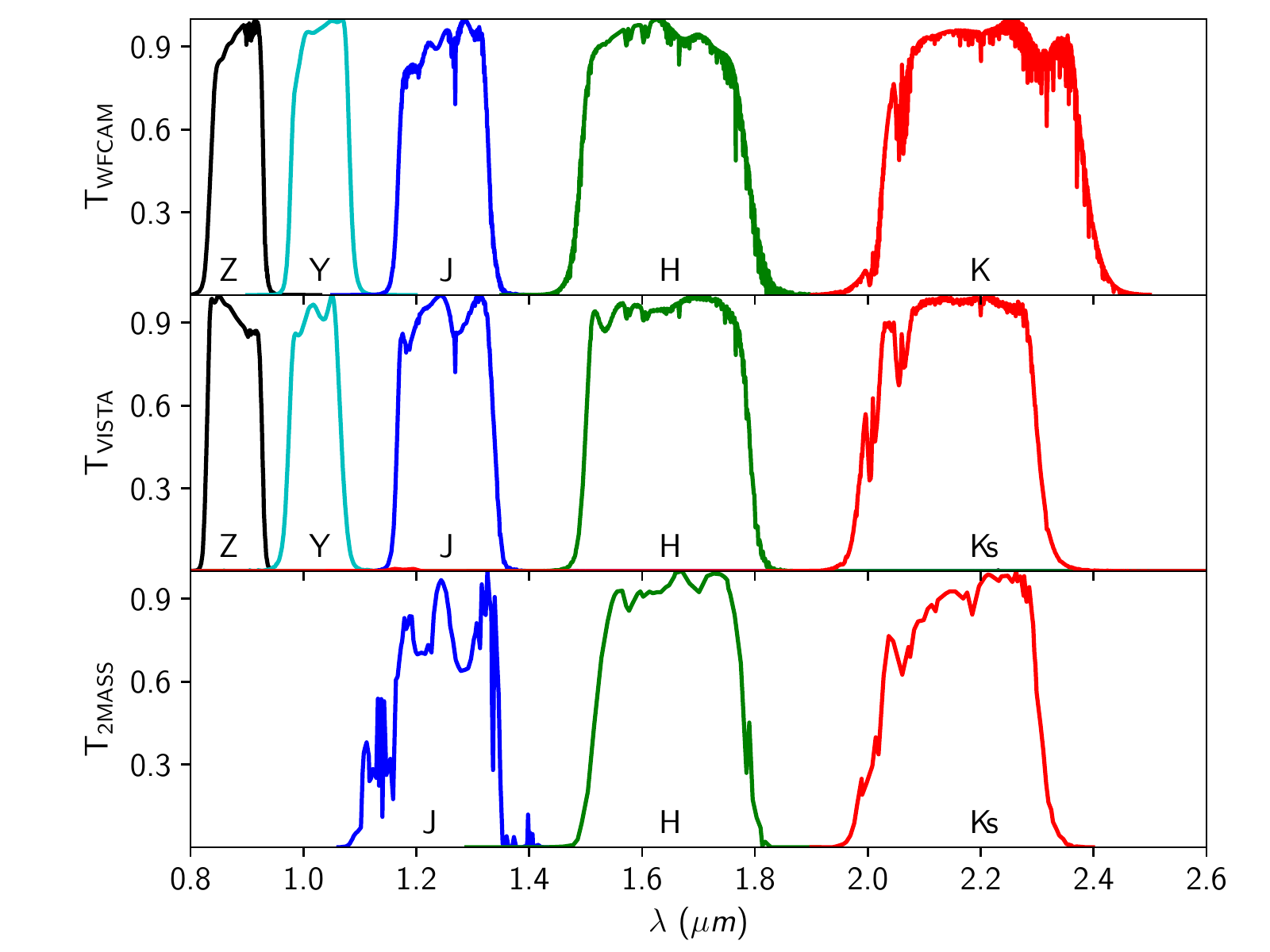}
 \caption{Relative spectral response curves (including the filter, the quantum efficiency of the detector and the effect of the atmosphere) of a typical detector-filter combination and atmosphere for 2MASS, VISTA and WFCAM. Every passband is normalized to its maximum value.}
\label{filt}
\end{center}
\end{figure}

The VISTA broadband photometric system is composed of 5 filters, ZYJHK$\mathrm{\!{s}}$ similar to those of WFCAM, with the exception of a shorter K band, closer to the one used by 2MASS (see Fig. \ref{filt}).

\subsection{VISTA instrumental magnitudes}\label{instru}
\label{insmag}
The source detection and extraction software \citep{mir85} measures an array of background-subtracted aperture fluxes for each detected source using 13 soft-edged circular apertures of radius $r/2$, $r/\sqrt{2}$, $r$, $\sqrt{2} r$, $2r$ ... up to $12r$, where $r=1$ arcsecond. A soft-edged aperture divides the flux in pixels lying across the aperture boundary in proportion to the pixel area enclosed. All the aperture fluxes of reliably morphologically classified stars are used to determine the curve-of-growth of the aperture fluxes, i.e. the enclosed counts as a function of radius. This curve of growth is used to measure the point spread function (PSF) aperture correction for point sources for each detector and for each aperture up to and including $4r$. Although an 8 arcsecond diameter aperture includes typically $\sim$99\,per cent, or more, of the total stellar flux, corrected aperture fluxes are also compared directly with isophotal fluxes corrected using a Moffat profile (VIRCAM images are well-described by a Moffat profile with $\beta = 2.5)$. Using field overlaps we find that this method derives aperture corrections which contribute $\leq1$\,per cent to the overall photometric error budget. In this paper we only consider photometry derived from fluxes measured within an aperture of radius 1 arcsecond, as it contains a reasonable fraction of the PSF (under typical seeing, $\sim75\%$ of the total flux) while being narrow enough that it performs well in crowded fields.

A further correction has to be applied to the source flux to account for the non-negligible field distortion in VISTA, described in detail in \citet{sut15} and \citet{irw04}. The astrometric distortion is radial and leads to a decrease in pixel sky area coverage by around 3.6\,per cent at the corners compared to the centre of the field-of-view. For ease of use and convenience standard flatfielding techniques assume a uniform pixel scale and hence that a correctly reduced image will have a flat background. For the variable pixel scale of VIRCAM this is clearly incorrect and one would expect to see a decrease in the sky counts per pixel at large off-axis angles. In contrast the total number of counts detected from a star would be independent of its position on the array. Conventional flatfielding of an image therefore introduces a systematic error into the photometry of discrete sources which varies as a function of distance $r$ from the optical axis and this error increases significantly towards the edge of the field-of-view. If the true angular distance is $r_{\rm true}$ then
\begin{equation}
r = \frac{1}{k_1} r_{\rm true} + \frac{k_3}{k_1} r_{\rm true}^3 + \frac{k_5}{k_1} r_{\rm true}^5 + \ldots
\end{equation}
where $k_1$ is the pixel scale at the centre of the field and $k_3$ and $k_5$ describe the angular radial distortion on the focal plane. For VISTA, $k_1$ = 0.3413 arcsec/pixel (i.e. 17.065 arcsec/mm) and in angular units in radians, the distortion coefficients are given by $k_3 = 44$ and $k_5 = -10300$. Higher order correction terms are negligible. Unlike WFCAM (see H09) there is no measurable dependence of the distortion terms with filter. 

\begin{figure}
\begin{center}
\includegraphics[width=9.0cm]{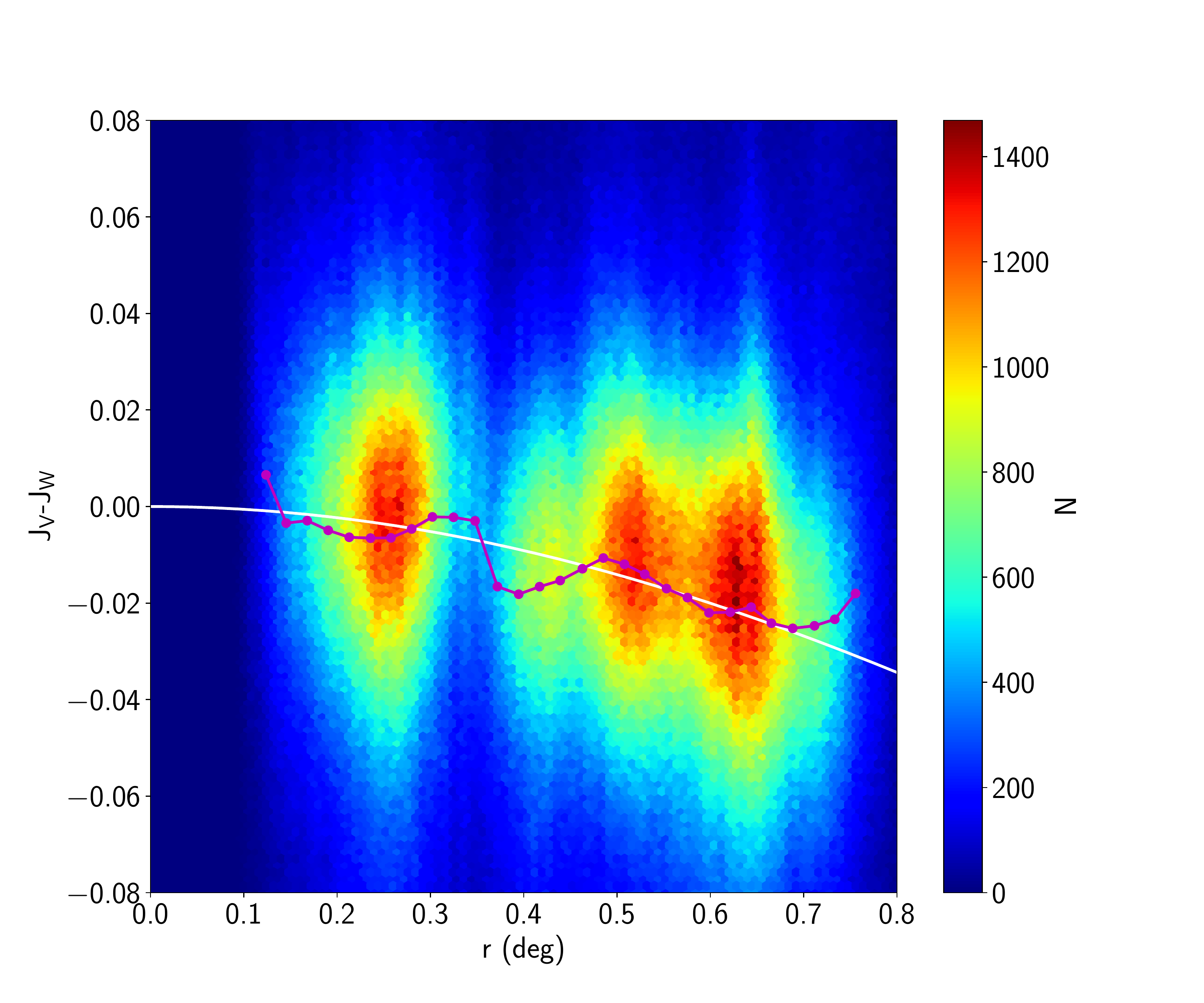}
\caption{Density plot of the magnitude difference between VISTA (uncorrected for radial distortion) and WFCAM as a function of off-axis angle (the result is the same using other surveys like 2MASS). Due to the radial distortion present in the VISTA system, pixels decrease in size with radius, so conventional flatfielding causes objects to appear brighter in the outer regions of the array. Red circles are a moving median, and the white line plots the predicted correction from Eq. \ref{eq:distortcor}. Higher order terms in the corners of each detector account for the see-saw pattern.}
\label{spatial}
\end{center}
\end{figure}

The corrected flux $f_{\rm cor}$, where $f$ is the measured aperture-corrected counts in ADU of the source above background, ignoring radial distortion terms of $O\left(r^7\right)$ and higher, is given by 
\begin{equation}
f_{\rm cor} = f / ( 1 + 3 k_3 r_{\rm true}^2 + 5 k_5 r_{\rm true}^4) 
( 1 + k_3 r_{\rm true}^2  + k_5 r_{\rm true}^4) .
\label{eq:distortcor}
\end{equation}
Fig.~\ref{spatial} illustrates the effect of this radial distortion on the photometry, as measured without correction for radial distortion comparing with WFCAM, and the suitability of the correction from Eq. \ref{eq:distortcor}.


\section{Combining Pawprints into tiles}

\begin{figure}
\begin{center}
\includegraphics[width=8.3cm]{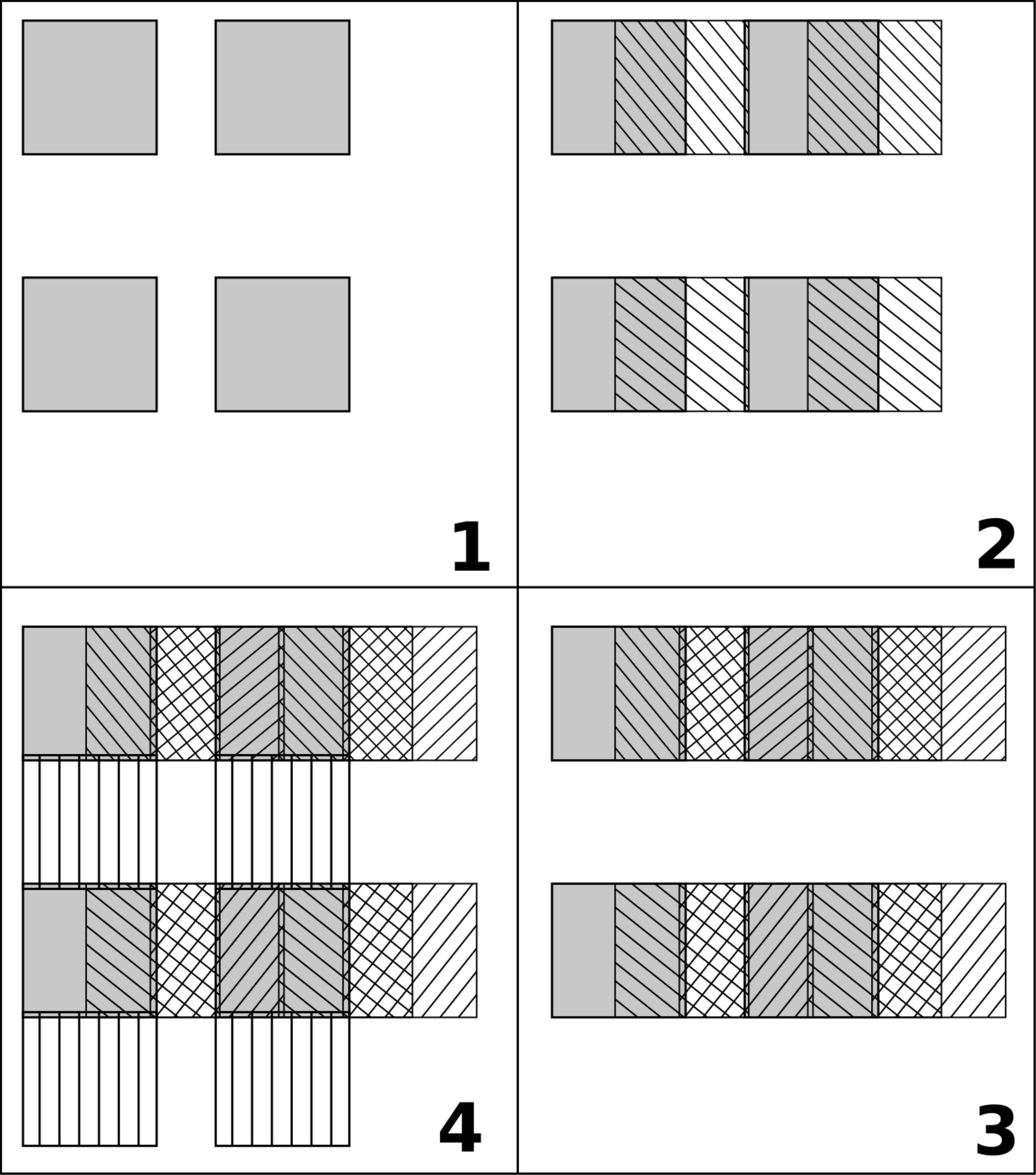}
\caption{Schema of the first 4 steps of a tile mosaic, only showing 4 detectors for clarity. Between 1 and 2, and 2 and 3 the telescope is offset in the X direction $47.5\%$ of a detector. Between 3 and 4 the offset in the Y direction is of $95.0\%$.}
\label{tileproc}
\end{center}
\end{figure}

\subsection{Observational strategy}
In order to get continuous coverage of the sky, the gaps between detectors seen in Fig. \ref{fplane} need to be filled. To do so, a minimum of 6 pointings are needed\footnote{ Several possible options are available for this pattern, they are described in more detail in the instrument pages
https://www.eso.org/sci/facilities/paranal/instruments/\\
vircam/doc.html}. This is the standard observing mode for VISTA and the pointings are distributed as follows: initially, three stacked pawprints are taken successively (although depending on the selected observing mode, not consecutively) with an offset $47.5\%$ of a detector in the X direction between them, as outlined in Fig. \ref{tileproc}. This produces three continuous strips of about 5.3 detector widths (or about 1.02$^\circ$). Once this is finished, the telescope is offset $95.0\%$ of a detector in the Y direction and the process is repeated again. After combining the 6 stacked pawprints, this results in a contiguous mosaic (tile) of $1.02^\circ\times1.47^\circ$, or about $12k\times16k$ pixels. All objects, except those in strips at the edge of the tile, are observed in at least 2 of the 6 paw prints comprising the tile.

In principle to create a tiled image all that is required is to adjust the sky level on all 96 component single detector images to the same level and then project (drizzle) the detector-level component parts onto a single (Tangent Plane WCS) image. During the projection, detector to detector differences are corrected, as each one will have its own measured zeropoint (Sect. \ref{comp2mass}). Pixel intensities are also adjusted with respect to sky, to account for the radial dependency on effective pixel size  introduced by the the distortion within the original ZPN WCS pawprints (eq. \ref{eq:distortcor}) and the much smaller photometric distortion $\leq 0.1$\% inherent in a Tangent Plane projection. Tile catalogues are then created directly from these mosaiced tile images.

\subsection{Complications of tiling}
Tiles contain 6 stacked pawprints each composed of 16 detector-level images. Therefore in practice there will potentially be 96 different sky levels present, 96 different PSFs, and in non-photometric conditions up to 6 different pawprint magnitude zeropoints. Although the flatfielding stage puts all detectors on a common internally consistent gain system based on analysis of twilight flats, gradients in the twilight sky, differences in the detector QE as a function of wavelength and the vagaries of NIR sky subtraction mean that variations in sky level between detectors are common. These issues affect both images and photometry catalogues, and therefore the correction for these detector to detector variations is applied to both data products. In contrast, all the corrections described in the next paragraphs and in Sect. \ref{groutsect} are only relevant to photometric catalogues and not applied to the delivered images themselves.

Each detector image that goes into a tile invariably has a different PSF due to slight differences in primary and secondary mirror settings. This together with the varying seeing conditions that happen during the observation means that there are potentially 96 different PSFs which contribute to a single tile. The potentially varying photometric throughput further complicates the issue.

In order to compute the correction for the aperture fluxes we normally approximate that the PSF does not vary across the image (tile or detector-level if pawprint). For a tile image this will inevitably introduce spatial photometric distortions at the level of a few per cent or more. In variable seeing conditions and/or variable throughput this effect will be significantly worse and can produce systematic spatial offsets 
of 10--20 per cent or more in the photometry. 

The first step in preparing the stacked pawprints for catalogue generation is to remove any major sky gradients on each detector and then compute a robust estimate of the sky level on every detector. Removing sky gradients is done using an iterative, clipped non-linear filter\footnote{The software is publicly available from http://casu.ast.cam.ac.uk and a description of the algorithm can be found in the Spring 2010 UKIRT newsletter available at http://www.ukirt.hawaii.edu/publications/newsletter/\\ukirtnewsletter2010spring.pdf}. Originally designed to track and filter out nebulosity in crowded star forming regions, this does an excellent job of removing sky background variations over a detector. This then greatly simplifies the task of enforcing a common sky level across a tiled image prior to cataloguing.

\subsection{Grouting}
\label{groutsect}

\begin{figure}
\begin{center}
\includegraphics[width=8.5cm]{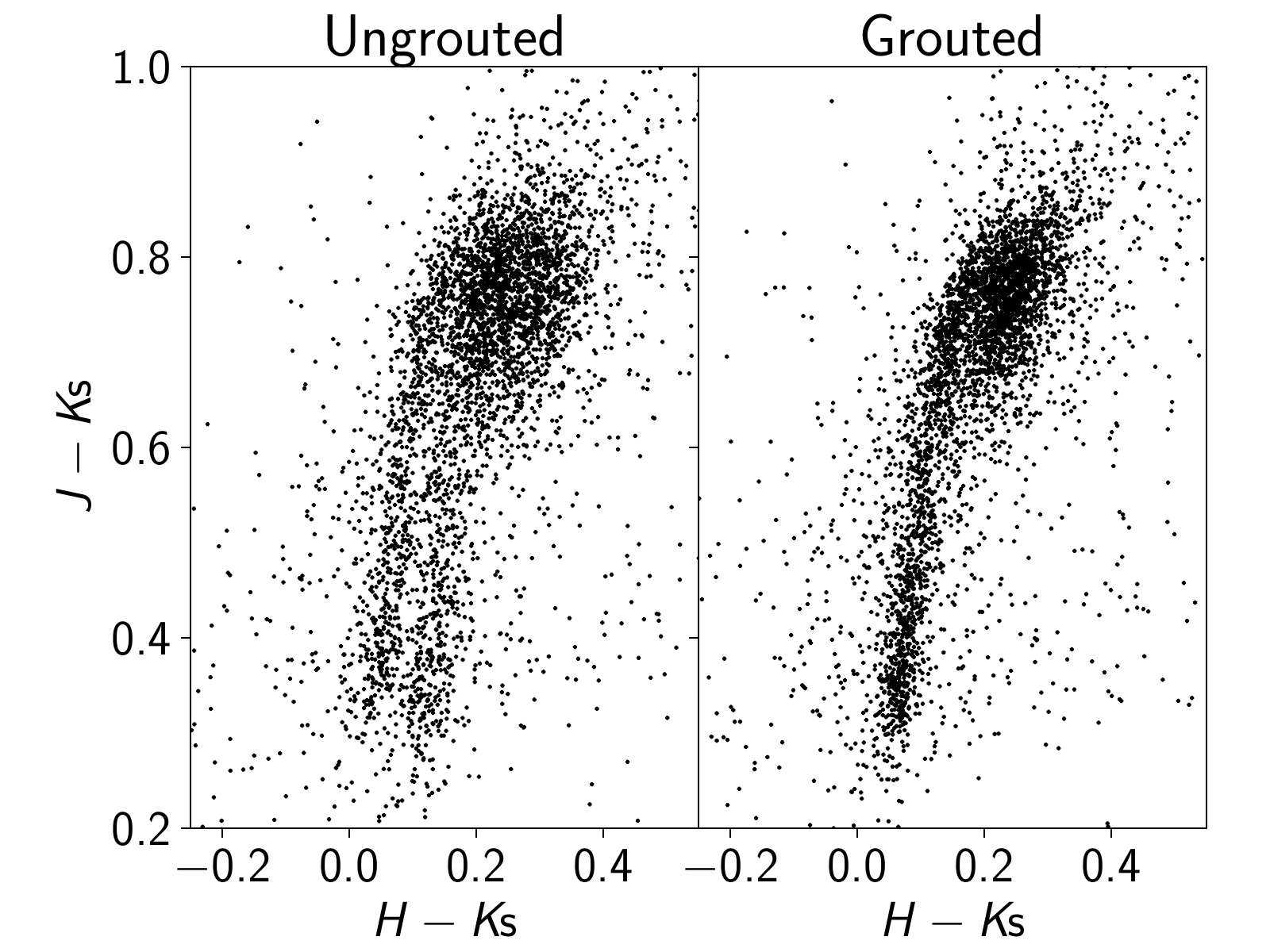}
\caption{Effect of the grouting correction on a colour-colour diagram of an off-plane tile. The uncorrected magnitudes (left) are affected by seeing changes between the different exposures that make the tile. This translates into a spatially variable aperture correction that has been taken into account on the right panel.}
\label{groutcmd}
\end{center}
\end{figure}

To counter the PSF variations we developed a piece of software to track and post-correct the variable flux within the main apertures for tile catalogues. The grouting fix takes as input the tile catalogue, the stacked pawprint catalogues and the associated confidence maps. These contain the ancillary information recording the individual aperture corrections and the monthly updates to the individual detector magnitude zeropoints (see section~\ref{detoff} for further details). Differential aperture corrections at the location of each detected object, weighted by the confidence map that was used to drive the tiling of the image, are then computed. The individual detector aperture corrections, 96 of them for each aperture, define the differential aperture corrections i.e. the difference with respect to the median for the whole tile. All fluxes and associated errors for apertures 1--7 are corrected, larger apertures are negligibly affected by seeing variations and are left unchanged. The updated catalogue then has to be re-classified and re-photometrically calibrated. The correction software optionally also adjusts for changes in the pawprint magnitude zeropoint, i.e. caused by atmospheric extinction variations, that occured during the pawprint observation sequences. The Observation Block (OB) for typical VISTA tiles takes between 10--60 minutes to execute hence variations in seeing and/or atmospheric transmission are not that uncommon, as can bee seen in Fig. \ref{groutcmd}. 
 
\section{Calibration of VISTA}
\label{comp2mass}
\subsection{Calibration strategy}
The strategy for VISTA follows that developed for the UKIRT WFCAM processing pipeline (H09). The procedure consists of these successive steps:
\begin{enumerate}
\item[1)] {\it Flatfielding and gain correction}: The data are flatfielded using twilight flatfields that are updated on a monthly timescale. Typically, VISTA takes one or two series of 15 twilight flats per night, each of which will be combined onto a single flat frame; in order to accumulate enough flats for each filter about four weeks are needed. Based on these flat frames, an initial gain correction is applied to place all sixteen detectors on a common system, at least to first order, by homogenizing the measured sky level in all the detectors.
\item[2)] {\it Zeropoint derivation}: The overall pawprint or tile magnitude zeropoint is then derived for each frame from measurements (in the VISTA photometric system) of stars in the 2MASS point source catalogue (PSC) that fall within the region observed. Thus the calibration stars are measured at the same time as the target sources. Observations of reference fields are also made typically at the start, middle and/or end of a night and treated in the same way.
\item[3)] {\it Detector offsets} The zeropoints derived from 2MASS use all the available stars in an image, but due to the flatfielding procedure (and other effects) each detector will have a slightly different response, that is corrected through an offset to its individual zeropoint on a monthly basis.
\item[4)] {\it Grouting}: Once pawprints are combined into tiles, one further calibration needs to be done. As tiled frames are composed of images taken over a span of time long enough that the observing conditions may change, this introduces a complicated PSF structure in the field that affects the fixed aperture photometry. This is compensated during the grouting process, using the zeropoints calculated in the previous steps to compute tailored corrections for each star in a tile. 
\end{enumerate}

\subsection{VISTA photometric system calibration using 2MASS stars}
\label{calto2mass}

The different system response functions (mainly detector + filter) between the VISTA and 2MASS systems require a transformation between the two photometric systems. We assume that for the majority of stars there exists a simple linear relation between the 2MASS and VISTA colours\footnote{Through these equations, we will use $(J-K\!s)_2$ as reference, although other colour combinations can be chosen, see Appendix \ref{ap:colour}.}, e.g. $J_{\rm V}-J_2 \propto (J-K\!s)_2$. In a Vega-based photometric system, this relation should pass through (0,0), i.e. for an A0 star, by definition  $Z$\,=\,$Y$\,=\,$J$\,=\,$H$\,=\,$K\!s$, irrespective of the filter system in use.

However, as discussed in section 4.4 the different effects of reddening in the different systems will also affect the transformation. For each star in 2MASS observed with VISTA in each filter, we therefore transform the 2MASS magnitude, $m_{\rm 2}$, into a magnitude in the VISTA system, $m_{\rm V}$, using
\begin{equation}
m_{\rm V} = m_{\rm 2} + C\cdot(J-K\!{\rm s})_2 + D\cdot E(B-V)
\label{eqn3}
\end{equation}
where $C$ is the colour coefficients for each pass-band determined by combining data from many nights (see Sections \ref{2mequations} and \ref{fitcolterm}), $D$ is the extinction coefficient at the appropriate band, (see Sect. \ref{intext} and Appendix \ref{cofext}) and E(B-V) is the B-V colour excess (see Section 4.4) as a measure of reddening. For each star in 2MASS observed with VISTA, we derive a ZP (at airmass unity) from
  
\begin{equation}
ZP  = m_{\rm V} + m_{\rm i} + k\cdot(\chi-1) 
\label{eqcol}
\end{equation}

where $m_{\rm V}$ is the relevant 2MASS magnitude put into the VISTA system, $m_i$ is the aperture-corrected physical instrumental magnitude i.e. $2.5\cdot log_{10}(flux/t_{\mathrm{exp}})$, $k$ is a default value for the atmospheric extinction (assumed to be $k=0.05$ in all filters, see below) and $\chi$ is the airmass. The zeropoints, zp, derived from all selected 2MASS stars in all 16 detectors are then combined to give a single zeropoint, ZP, for that pawprint.

To derive the calibration zeropoint for an individual VIRCAM detector, the 2MASS sources are selected according to the following prescription:

\begin{enumerate}\itemsep6pt
\item{We use only VIRCAM detections morphologically classified as unsaturated stellar objects, as VIRCAM has a higher spatial resolution. These will generally correspond to point-like sources in 2MASS, with the exception of close groups that may be unresolved at lower resolution and that will introduce an asymmetry in the tails of the magnitude difference distribution.}
\item{These sources are matched against the 2MASS Point Source Catalogue (PSC) with a maximum allowed separation of 1 arcsecond. The VIRCAM astrometric calibration is also derived from the 2MASS PSC which means that in practice differential astrometric misalignment is dominated by the 2MASS source $rms$ precision which for matched source averages $<$100mas.}

\item{Next we restrict the calibration set to those objects with 2MASS $(J-K\!{\rm s}) \leq 2$, an extinction-corrected (see Appendix \ref{cofext}) 2MASS colour in the range $0.0 \le (J-K\!{\rm s})_{0} \le 1.0$ and a 2MASS signal-to-noise ratio $>10$ in each passband, the latter criterion minimises the effects of Malmquist bias at faint 2MASS magnitudes. Then:
\begin{itemize}
\item[a)]{If fewer than 25 (800) 2MASS sources remain within the field-of-view per detector (tile) in the calibration set, then the 2MASS $(J-K\!{\rm s})$ colour cut is not applied but the extinction-corrected 2MASS colour cut $0.0 \le J-K\!{\rm s}_{\rm o} \le 1.0$ is still used.}
\item[b)]{If this still results in fewer than 25 (800) 2MASS sources per detector (tile) no colour cuts are applied.}
\end{itemize}}
\end{enumerate}

\begin{figure}
\begin{center}
\includegraphics[width=8.5cm]{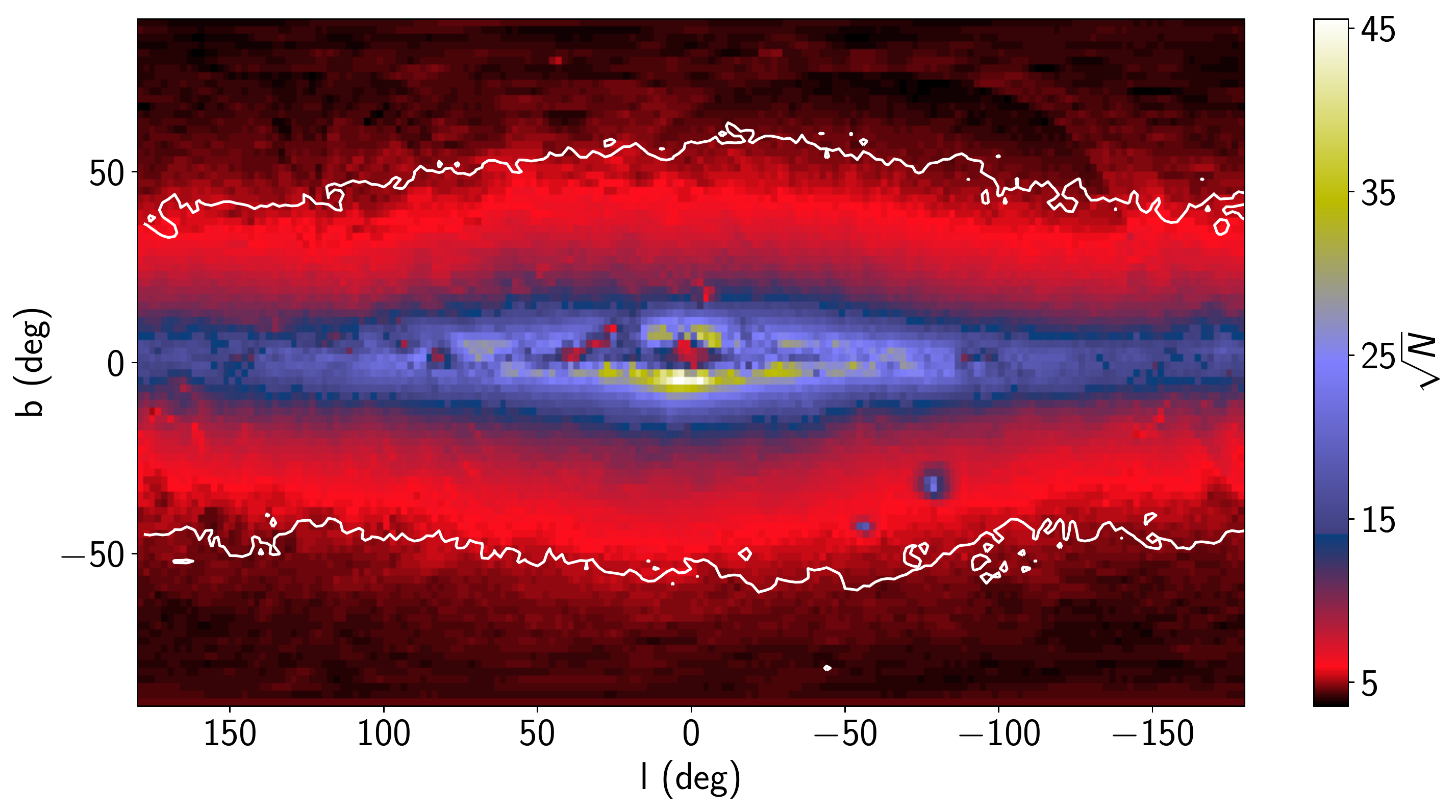}
\caption{Distribution of the available sources, per detector, from 2MASS for the derivation of the zeropoints. The white contour marks $N=25$. For the selection criteria, see Sect. \ref{comp2mass}.}
\label{ndens}
\end{center}
\end{figure}

In Fig. \ref{ndens} we show the 2D distribution of available 2MASS sources following the first set of aforementioned colour selection criteria. The density falls below 25 sources/detector only beyond $|b|\sim50^\circ$. It is worth emphasising that zeropoints are derived per pawprint and not per detector, so good sample numbers are available even in areas with a very low density of calibrators; in particular, pawprint stacks rarely have less than 100--200 a priori possible calibrators. As we require that these stars have photometric errors in 2MASS below $10\%$, it is possible a priori to determine the zero point of a pawprint to $1\%$ or better. Rather than modifying the measured fluxes, these zeropoints are encoded into the headers of images and catalogues. A brief description of the of these headers and the relevant keywords can be found in Appendix \ref{headers}.

\subsection{2MASS to VISTA Colour Equations}
\label{2mequations}

As the $JHK\!{\rm s}$ filter set for VISTA differs to that used by 2MASS, colour terms arise when comparing both photometric systems. Furthermore, there is no equivalent filter for $Y$ and $Z$, so in order to apply Eq.~\ref{eqcol}, we need a robust calibration of the colour terms that will enable the 2MASS to VISTA transformations. For a given passband and we can use the wealth of observations already taken by VISTA to robustly derive the desired colour terms.

An additional complication arises due to the different system response functions (mainly detector $+$ filter) between the photometric systems. In addition to the usual colour terms, an extra correction will be required due to the differential effects of interstellar extinction (reddening) between the VISTA and 2MASS systems. We defer a fuller discussion of this until the next section (section~\ref{intext}).

As Eq.~\ref{eqn3} shows, to transform 2MASS magnitudes to the VISTA system we need to determine the colour coefficients, C, and the extinction correction coefficients, D for each VISTA filter. We defer a fuller discussion of determining D until section~\ref{intext}, and we proceed to calculate the colour coefficients using stars measured in nights with good photometric quality (as parametrized by the bandpass zeropoint variances during the night) at high latitude fields ($|b|\geq35^\circ$) that satisfy $E(B-V)\leq0.1$ \citep[from][]{sche98}. This way we can neglect the extra extinction correction while guaranteeing that the introduced error per star will be less than 1\%. To further minimize systematic errors, we require the 2MASS stars to have signal-to-noise ratio $>$10 in each 2MASS filter and to be morphologically classified as unsaturated stellar objects in the VISTA data.This leaves about $10^4$ to $10^5$ stars per month, depending on the filter (Z being the least used and K\!s the most).

These colour terms were derived from linear fits to the 2MASS-VISTA magnitude difference with respect to $(J-K\!s)$. For convenience, the theoretical, zeropoint of the system\footnote{ Determined from the transmission curves of the optical system and filters, and the quantum efficiency of the detector} is added to $m_\mathrm{i}$ so that the magnitude differences are close to zero.

As will be shown later (in sections \ref{detoff} and \ref{longterm}), the actual zeropoint of the system changes with time and detector, while we effectively assume a single value per filter. To minimize the dispersion that this simplification introduces into the data, we do the colour equation fits on a monthly basis. By restricting the time interval we also minimize possible seasonal variations while still guaranteeing that there are sufficient reddening-free stars to provide a good fit.

A month is also the time scale on which we update the flatfields (and therefore the per-detector zeropoint variations). As there is no evidence for significant temporal variation (sect. \ref{longterm}), from the derived monthly values of the colour coefficients we calculate averages and deviations of the colour terms weighted with the number of observations taken per month.

\begin{figure*}
\begin{center}
\includegraphics[width=18cm]{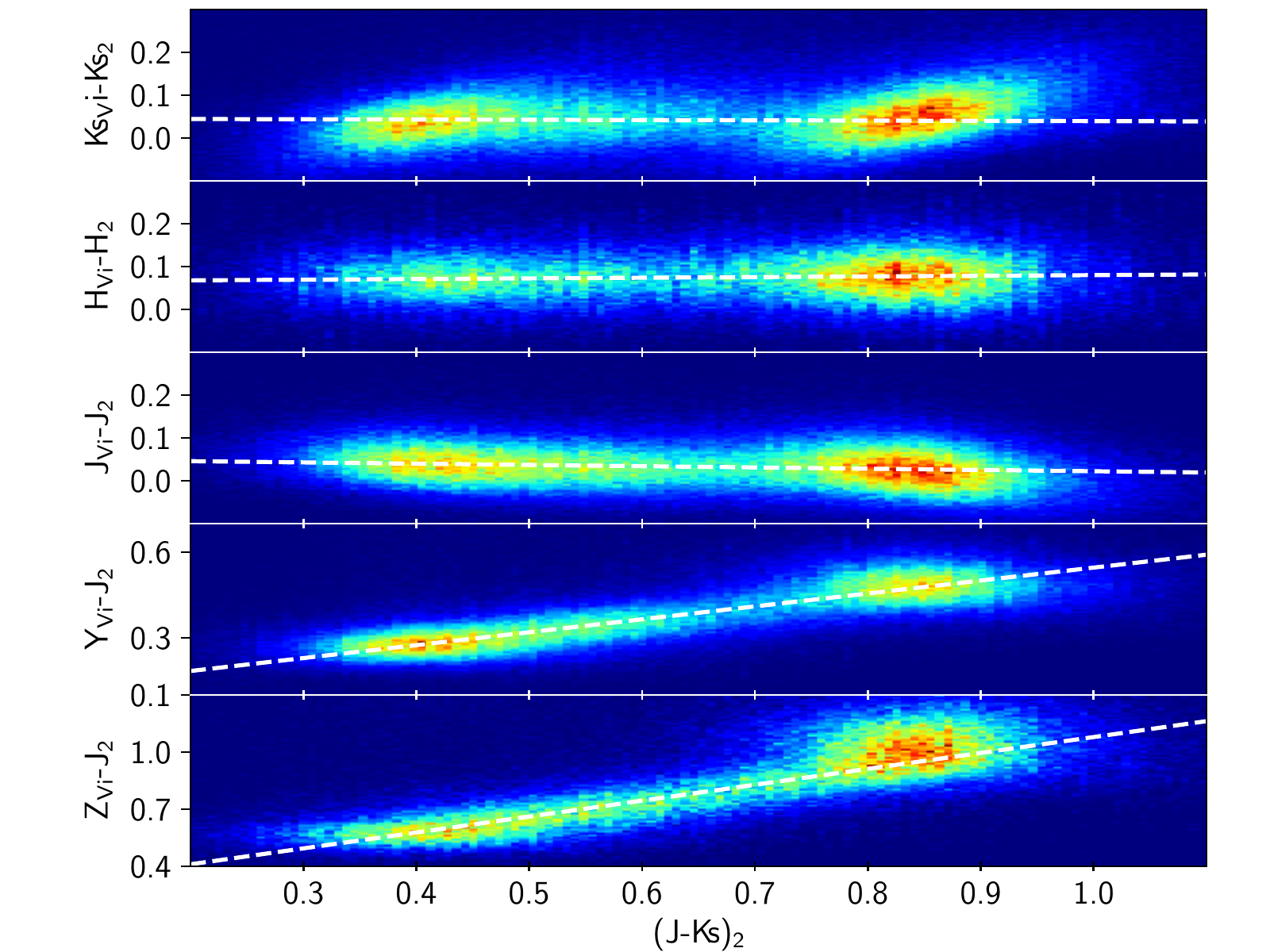}
\caption{A Hess diagram showing differences between VISTA and 2MASS  photometry as a function of 2MASS $(J-K\!{\rm s})_{\rm 2}$ (Eqs. \ref{eq:zj} to \ref{eq:ks}), for data taken during 2015. The best-fitting linear regressions to the unbinned data are overplotted. {\bf VISTA magnitudes are instumental, therefore the $i$ subindex.}}
\label{fig:colourequations}
\end{center}
\end{figure*}

Fig.~\ref{fig:colourequations} shows an example of this for data taken throughout 2015, whilst Eqs. \ref{eq:zj} to \ref{eq:ks} show the results from this procedure, averaging values taken during the entire 2010-2015 period. The fitting algorithm and colour choices are detailed in Appendix \ref{ap:colour}.

\begin{align}
Z_{\rm V} = J_{\rm 2} + (0.86\pm0.08)\cdot(J-K\!{\rm s})_{\rm 2}\label{eq:zj}\\
Y_{\rm V} = J_{\rm 2} + (0.46\pm0.02)\cdot(J-K\!{\rm s})_{\rm 2}\label{eq:yj}\\
J_{\rm V} = J_{\rm 2} - (0.031\pm0.006)\cdot(J-K\!{\rm s})_{\rm 2}\\
H_{\rm V} = H_{\rm 2} + (0.015\pm0.005)\cdot(J-K\!{\rm s})_{\rm 2}\\
K\!{\rm s_V} = K\!{\rm s_2} - (0.006\pm0.007)\cdot(J-K\!{\rm s})_{\rm 2} \label{eq:ks}
\end{align}

\subsection{Allowing for interstellar extinction}\label{intext}

The VISTA calibration makes two basic assumptions. Firstly, that on average the population of calibrator stars observed in a VISTA pawprint is a representative sample wherever we look. Secondly, that the effects of interstellar extinction can be treated with a simple model.

In H09, we demonstrated that Galactic extinction has a significant effect on the calibration of WFCAM, especially for the $Z-$band. H09 presented a simple correction to the WFCAM calibration, which included a linear term dependent on a modified\footnote{using the recipe from \citet{bon00}} E(B-V) derived from the \citet{sche98} extinction maps evaluated at the position of each calibrator star. For the WFCAM $YJHK$ bands regions with E(B-V)$<1.5$, the resulting calibration errors are better than 2 per cent and thus meet the goals of the UKIDSS surveys. For the WFCAM Z-band, the scatter in the calibration is more like 3--4 per cent above E(B-V)=0.2. H09 note that some of this problem is caused by the effects of a paucity of $ZY$ photometry at regions of significant E(B-V) (the UKIDSS Galactic Plane Survey includes only JHK filters \citep{luc08}, and a lack of measurements in all filters for regions with very high reddening, E(B-V)$>$3.

It is useful to illustrate the effects of extinction using the $Y$ calibration as an example. For a given 2MASS calibration star with no extinction, we have determined that:
\begin{equation}
Y_V = J_2 + C_Y\cdot(J-K\!s)_2 .
\end{equation}
However, normally, all the stars we observe suffer some interstellar reddening hence we really measure reddened magnitudes, denoted by $Y'_V$, $J'_2$ and $K'\!s_2$. The colour relation is calibrated at effectively zero extinction, so we need to estimate $Y_V$, $J_2$ and $K\!s_2$.  If we know the line-of-sight extinction to the star and $R_\mathrm{YV}$, $R_\mathrm{J2}$, $R_\mathrm{K\!s2}$ the ratios between $E(B-V)$ and the extinction in 
the $Y$, $J_2$ and $K\!s_2$ passbands then 
\begin{align}
Y'_V - R_\mathrm{YV}\cdot E(B-V) = J'_2 + C_\mathrm{Y}\cdot(J'-K'\!s)_2 -\\ 
\left[R_\mathrm{J2}+C_Y\cdot(R_\mathrm{J2}-R_\mathrm{K\!s2})\right]\cdot E(B-V) \nonumber
\end{align}
and hence
\begin{equation}
Y'_V = J'_2 + C_\mathrm{Y}\cdot(J'-K'\!s)_2 + D_\mathrm{Y}\cdot E(B-V)
\end{equation}
with
\begin{equation}
D_\mathrm{Y}=\left[R_\mathrm{YV} - R_\mathrm{J2} - C_\mathrm{Y}(R_\mathrm{J2} - R_\mathrm{K\!s2})\right]
\label{eqford}
\end{equation}
  i.e. to first order we expect a linear dependence with extinction of the transformation from the 2MASS to VISTA photometric systems, as assumed previously in eq. \ref{eqn3}. Although in theory we could calculate all the D coefficients using eq. \ref{eqford}, in practice, due to uncertainties/variations in Galactic extinction laws, the values for these coefficients need to be refined, as it is explained in Appendix \ref{cofext}.

The initial estimate of $E(B-V)$ is derived from \cite{sche98}, interpolating from the closest 4 pixels to any star. Following Appendix \ref{cofext}, we convert this to the expected extinction values in the relevant passbands. As \cite{sche98} is nominally the total line-of-sight extinction through and even out of the Galaxy, in many directions close to the Galactic plane this results in an overestimate of the extinction to the star sometimes leading to negative values of extinction-corrected $(J-K\!s)_0$.  To counter this, in all cases where the \cite{sche98} extinction-corrected $(J-K\!s)_0$ is $\leq$0.25 we recompute the extinction to the star by forcing the extinction-corrected colour of the star to be $(J-K\!s)_0=0.5$. Over the field-of-view covered by either a pawprint or a tile we are assuming that variations in the encountered individual stellar types average out.

\begin{figure*}
\begin{center}
\includegraphics[width=18cm]{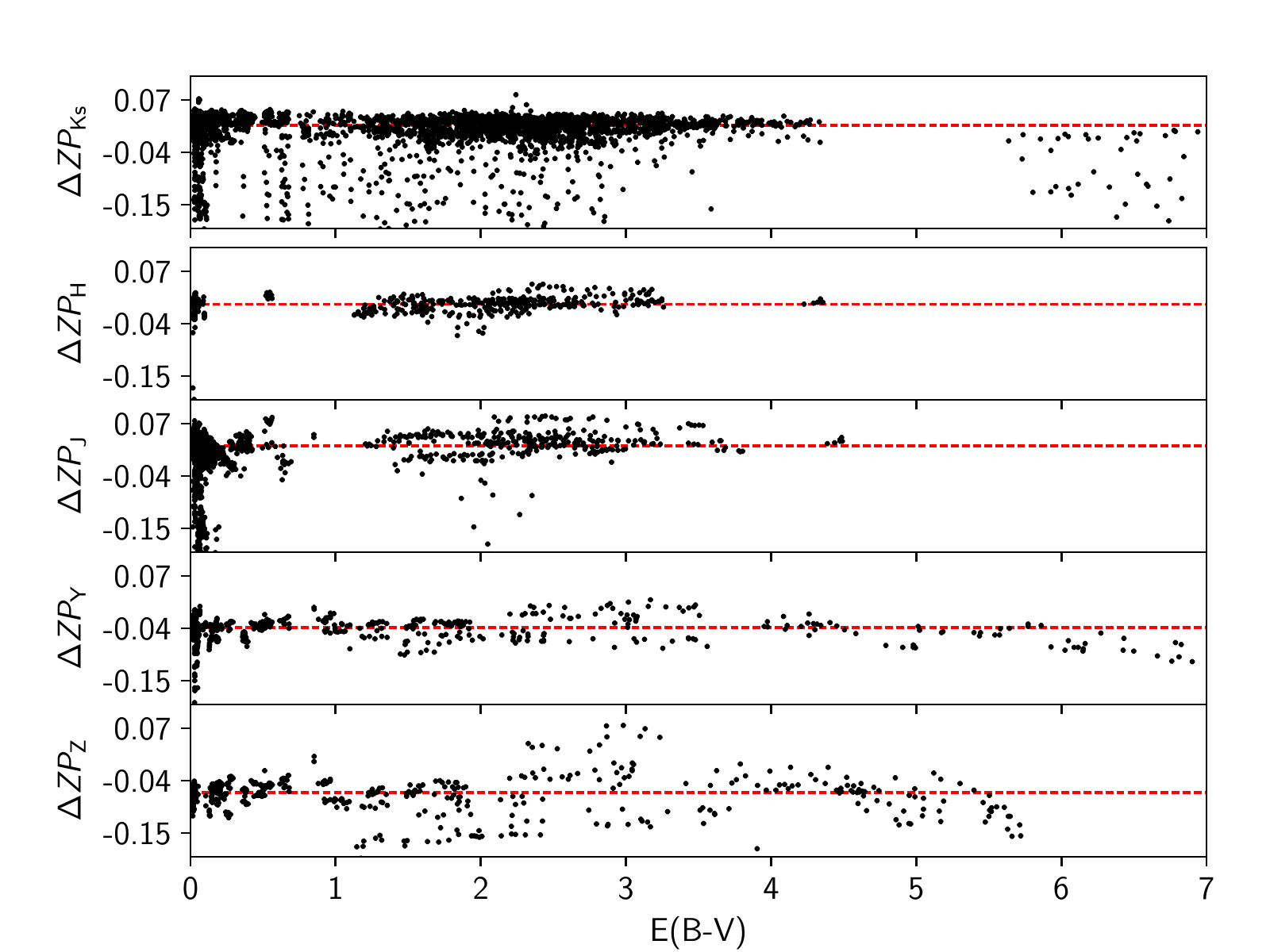}
\caption{ZP variation, expressed as the difference between the calculated ZP with respect to the theoretical ZP of the passband, as a function of $E(B-V)$. These values are computed for May 2015, where suitable fields at a wide range of Galactic latitudes were observed. No filter has been applied to the photometry, so the effect of atmospheric transmission variations from night to night are still visible.}
\label{zpebv}
\end{center}
\end{figure*}

As an illustration of this process in Fig. \ref{zpebv} we show the variation in computed zeropoints,  using the colour terms from eqs. \ref{eq:zj} to \ref{eq:ks} and the aforementioned recipe for extinction, as a function of the computed average (median) line-of-sight $E(B-V)$. The data used here were taken in May/June 2015 in a series of fields covering a large range of extinction both in and out of the Galactic Plane. As can be seen, for most values of $E(B-V)$ there is no residual dependency between the derived ZP and reddening in any of the filters. For fields with $E(B-V)>5)$, there is still some uncorrected colour term, very likely to be related to a change in the shape of the extinction law, as these fields are likely to be sampling dense molecular material close to the Galactic plane. 

\subsection{Initial Detector Zeropoints}
\label{inizero}

For each stacked pawprint a single overall zeropoint, MAGZPT, is derived as the median of all the per-star zeropoint values determined using eqn. \ref{eqcol}, and with the the relevant colour term and extinction coefficient described in previous sections. The associated error, MAGZRR is computed using the variation in the individual detector zeropoints using 1.48$\times$ the median absolute deviation (MAD) of the detectors zeropoints as the estimator. For tiles we do something similar to derive the MAGZPT except here the error (MAGZRR) is derived from the MAD estimator of the variation in the
pawprint MAGZPTs.

The error in an overall zeropoint has several contributors: intrinsic $rms$ and systematic errors in the 2MASS photometry; errors in the conversion from 2MASS to the VIRCAM filter set; errors in the determination of the interstellar extinction terms; intrinsic systematic offsets between the detectors; variations in atmospheric extinction; errors in the aperture corrections and so on. 

All estimates of the overall error based on the $1/\sqrt{n}$ weighted $rms$ scatter of the individual contributing stars severely underestimate the zeropoint error due to the large number of stars used in the solution. To counter this, and to include a more realistic assessment of some of the systematic errors noted previously, we find in practice that the variation (from the MAD) of the individual detector zeropoints provides a more reliable error estimate particularly in regions of high extinction. In the $Z$-band, the MAGZRR errors are the largest, typically about twice that for the $J$$H$$K\!{\rm s}$ filters with the $Y-$band in between.

It should be noted that we do not derive any atmospheric extinction terms on a given night with VIRCAM. Rather, the value of MAGZPT derived above incorporates an instantaneous measure of extinction at the observed airmass. The photometric calibration of a field therefore includes no error from this assumption. However the derived zeropoint normalised to airmass unity will include a small error, as we assume the extinction is 0.05 magnitudes per airmass for all filters \citep[this value is a good compromise in this wavelength range,][]{lom11}. For a typical VIRCAM frame, observed at an airmass $x\approx 1.3$, an extinction which differs from our assumed value by 0.03 magnitude/airmass would lead to $\leq$0.01 magnitude offset in the value of MAGZPT extrapolated to airmass unity. 

The value of MAGZPT over time can be used to investigate the long term sensitivity of VIRCAM due to, for example, the accumulation of dust on the optical surfaces, and seasonal variations in extinction. We explore this in more detail in Section \ref{longterm}.

\subsection{Detector offsets}
\label{detoff}

The final stage to the pawprint photometric calibration takes account of systematic differences between the 16 detectors, measured on a monthly basis. As every month we derive a new set of static calibration files, this step effectively homogenizes the flat field calibration from month to month. At the end of each monthly cycle, an illumination correction for the VISTA stacked pawprints is computed from the residuals between the VISTA magnitudes and the 2MASS magnitudes converted to the VISTA system. These illumination correction maps are primarily used to compute differential detector zeropoint offset which are then used to update the individual detector zeropoints for the month. This latter step is required before the grouted VISTA tile catalogues are generated. The illumination correction maps are computed on a rotated to PA$=0$ $\xi,\eta$ standard coordinate grid relative to the telescope pointing (equivalent to a focal plane X-Y system) which both allows for the different rotation angles of the observations and allows for the stars to be accurately located on the focal plane via the detector-level WCS solutions. Typically, for any given detector the variation on these coefficients is about $\pm1\%$, while the detector-to-detector amplitude is about $2\%$. There is no evidence of any temporal correlation.  

The pipeline also estimates the nightly zeropoint (NIGHTZPT) and an associated error (NIGHTZRR) which can be used to gauge the photometricity of a night. NIGHTZPT is simply the median of all ZPs for a given passband measured within the night, while NIGHTZRR is a measure of the scatter in the NIGHTZPT.

\section{Monitoring the long term sensitivity of VISTA}
\label{longterm}

For every image taken by VISTA and processed by CASU we store original and derived FITS header keywords in a local database. This makes, for example, an investigation of the distribution of QC (Quality Control) parameters fairly straightforward, and enables us to select good quality data measured on clear nights.

For the purposes of this paper, we define good quality data as the 75th percentile across a number of parameters, as listed in Table~\ref{tab:qcparams}. 

\begin{table*}
\centering
\caption{A sample of VISTA Quality Control parameters, and median values. For each filter we also give the values at the 75th percentile, in the sense that 75\% of the data have a value that is lower than the value in the table.}
\begin{tabular}{| l | rr | rr | rr | rr | rr |}\hline
 & \multicolumn{2}{c}{Z} & \multicolumn{2}{c}{Y} &
 \multicolumn{2}{c}{J} & \multicolumn{2}{c}{H} &
 \multicolumn{2}{c}{Ks} \\
 & 75\%ile  & 50\%ile & 75\%ile  & 50\%ile & 75\%ile  &
 50\%ile & 75\%ile  & 50\%ile & 75\%ile & 50\%ile\\ \hline
Seeing     & 1.046   & 0.889 & 1.126 & 0.912 & 1.064 & 0.868 & 1.012 & 0.840 & 0.941 & 0.784\\
Ellipticity     & 0.079   & 0.062 & 0.074 & 0.059 & 0.078 & 0.062 & 0.079 & 0.061 & 0.099 & 0.074\\
MAGZERR     & 0.063   & 0.045 & 0.031 & 0.024 & 0.019 & 0.014 & 0.027 & 0.021 & 0.022 & 0.017\\
NIGHTZRR     & 0.091   & 0.063 & 0.035 & 0.023 & 0.027 & 0.017 & 0.019 & 0.012 & 0.013 & 0.010\\
SKYBRIGHT*     & 18.1   & 18.4 & 16.8 & 17.1 & 15.5 & 15.8 & 13.6 & 13.9 & 12.8 & 12.9\\
STDCRMS     & 0.154   & 0.102 & 0.157 & 0.129 & 0.148 & 0.117 & 0.154 & 0.126 & 0.104 & 0.078\\
\hline
\end{tabular}
\begin{flushleft}
\small{*In this case, $75\%$ and $50\%$ of values are brighter, i.e. have lower magnitude.}
\end{flushleft}
\label{tab:qcparams}
\end{table*}

\begin{figure*}
\begin{center}
\includegraphics[width=18cm]{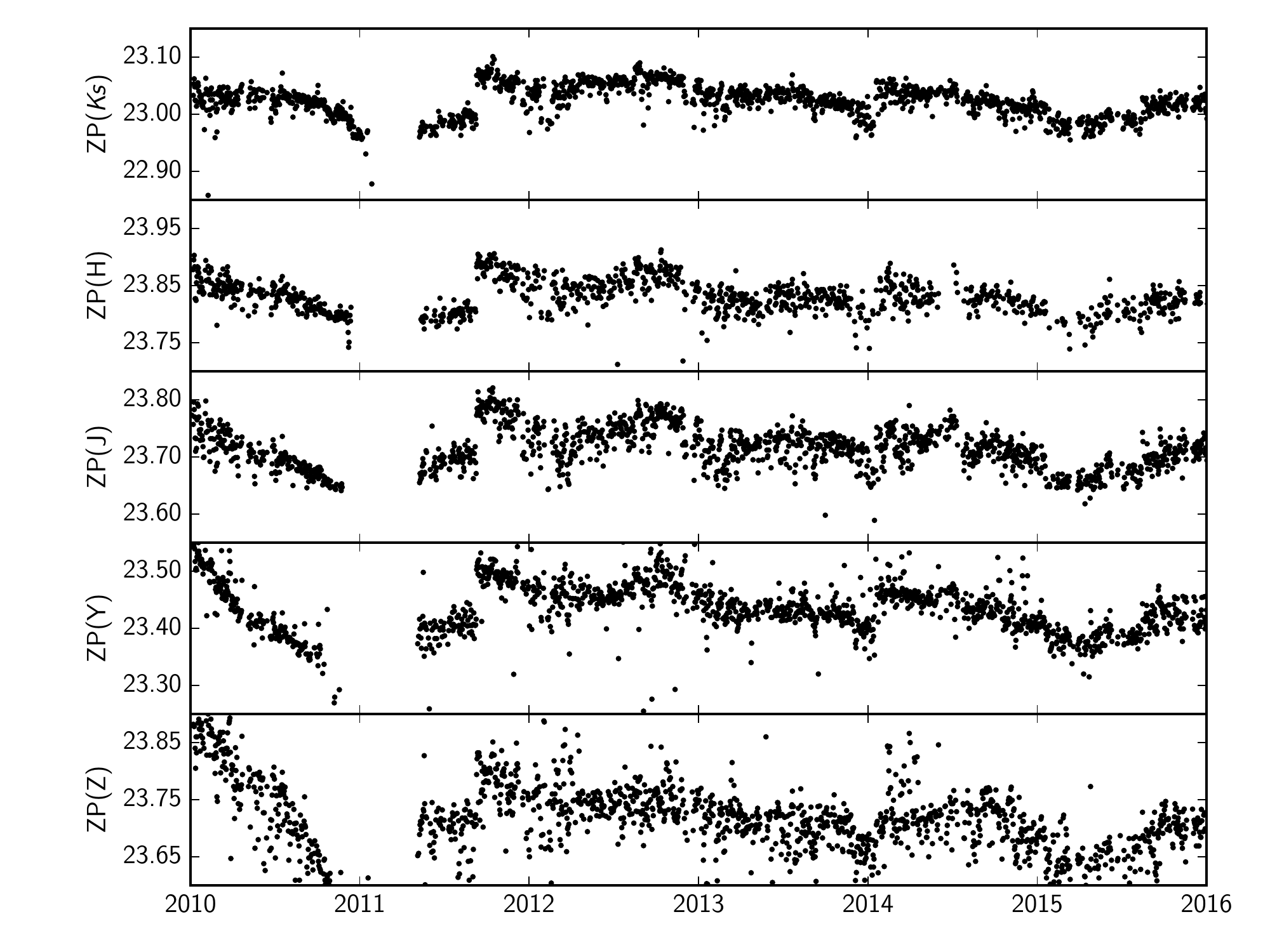}
\caption{Temporal evolution of the zeropoints, as measured from stacked images taken in good nights (i.e. NIGHTZRR below the 75th percentile).}
\label{hZP}
\end{center}
\end{figure*}

With a selection of good quality observations, we can monitor the temporal evolution of the photometric system. One way to do so is by checking the evolution of the zeropoints per filter, as they will reflect the overall sensitivity of the system, including the atmosphere at Paranal. This can be seen in Fig. \ref{hZP}. A number of clear features stand out: (i) The early slow decline in sensitivity in all passbands, but strongest in the bluer filters. (ii) A $\sim 2$ month gap in operations while the mirror was recoated (with Aluminium to replace the original Silver coating) and an extended repair following the discovery of a mechanical problem. (iii) Jumps in sensitivity on 09/2011 and 02/2014 caused by cleaning the VIRCAM entrance window (which also lowered the background). Some outliers are also evident. While points that fall below the main curve correspond to nights with worse than average transmission, those that show zeropoints larger than average normally come from observations of fields under very high interstellar reddening, where the algorithm to de-redden 2MASS sources fail. These are much more obvious in $Y$ and $Z$, where the dependence on interstellar extinction is much stronger.

These changes in sensitivity, and other effects, may also reflect on the colour terms derived in Sec. \ref{2mequations}. We can check this by looking at the variation of the colour terms from Sect. \ref{2mequations} with time. While the changes fall mostly below the typical error for a single measurement, there seems to be some minor seasonal trend. This is in part explained by the fact that while winter months (Southern Hemisphere) tend to be dominated by Galactic plane observations, in summer time off-plane regions dominate. This implies a difference on the typical stellar population used to derive the colour term, that could affect slightly the derived value. In any case, the existence of a temporal modulation of the colour term is weak, of about $1.5\sigma$, and therefore we opt to use a single averaged value per filter.


\section{Comparison with the UKIRT WFCAM Photometric System}
\label{vistavwfcam}

Both WFCAM and VISTA repeatedly target a set of reference fields, some of which are roughly equatorial and are observed by both telescopes routinely. Both instruments also have overlapping observations of off-plane fields. We select reddening-free stars (with $E(B-V)\leq0.1$) with good photometry, so that the errors in colour and magnitude difference are below 0.1 mag (this effectively clips out non-photometric nights, as they will have large ZP dispersion). These datasets can be used to cross-calibrate both photometric systems.

\begin{figure*}
\begin{center}
 \includegraphics[width=18cm]{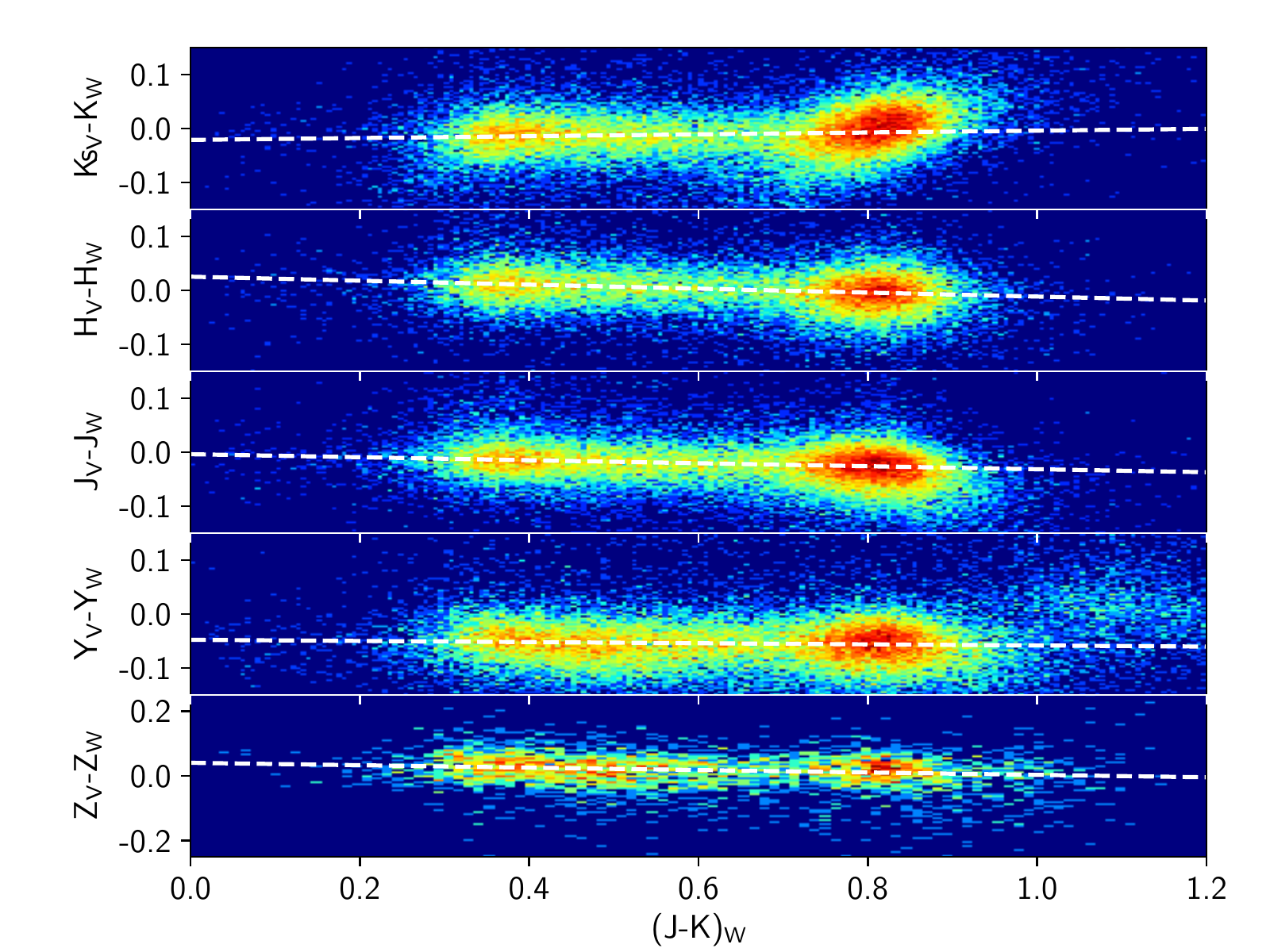}
 \caption{Magnitude differences between VISTA and WFCAM ($m_{\rm VISTA}-m_{\rm WFCAM}$ as a function of WFCAM $J-K$ colour for the $ZYJHK(s)$ filters. The dashed lines mark the fits from Eqs. \ref{eq:wzeq} to \ref{eq:wkeq}.}
\label{viwf}
\end{center}
\end{figure*}

In figure~\ref{viwf} we show the differences between the VISTA and WFCAM photometry for stars in these comparison fields. Small colour terms and offsets are apparent between the two photometric systems. Recall that, even though both systems depend on the same photometric calibrators from 2MASS, their photometry is converted into the native system of the specific instrument. WFCAM and VISTA are thus on different photometric systems, and such terms and offsets are expected.

The transformations between VISTA and WFCAM are described in equations~\ref{eq:wzeq} through~\ref{eq:wkeq} and are derived from robust linear fits to the data presented in figure~\ref{viwf}. Errors in the coefficients are obtained through 1\,000 Monte-Carlo fits, in which we substitute each measured star by a normal distribution with its colour and magnitude difference as $\mu$ and the respective errors as $\sigma$. The colour terms are all small but significant, as expected due to the fact that both filter sets are similar. There are also statistically significant offsets that will be discussed in Sect. \ref{overper}.

\begin{equation}
Z_{\rm V} - Z_{\rm W} = - (0.037\pm0.008)\cdot(J-K)_{\rm W} + (0.040\pm0.005)
\label{eq:wzeq}
\end{equation}

\begin{equation}
Y_{\rm V} - Y_{\rm W} = - (0.010\pm0.003)\cdot(J-K)_{\rm W} - (0.048\pm0.002)
\label{eq:wyeq}
\end{equation}

\begin{equation}
J_{\rm V} - J_{\rm W} = - (0.028\pm0.002)\cdot(J-K)_{\rm W} - (0.004\pm0.001)
\label{eq:wjeq}
\end{equation}

\begin{equation}
H_{\rm V} - H_{\rm W} = - (0.037\pm0.001)\cdot(J-K)_{\rm W}+(0.025\pm0.001)
\label{eq:wheq}
\end{equation}

\begin{equation}
K\!s_{\rm V} - K_{\rm W} = (0.017\pm0.003)\cdot(J-K)_{\rm W} - (0.022\pm0.002)
\label{eq:wkeq}
\end{equation}

\section{Overall performance of the pipeline}
\label{overper}
\subsection{Photometric calibration}
\subsubsection{Internal consistency}
\label{phores}
The final uncertainty in the delivered photometry has two main components. One is systematic and has to do with the absolute calibration of the photometric system, while the other is random and is related to both the photometric measurement itself, particularly for stars close to the sensitivity limit or out of the linear response regime of the detector, and with the error in the determination of the zeropoint of a given image. Being random, we can characterize it by looking at repeated measurements of a field; reference fields, as they are observed with a high cadence, are the obvious choice for this. By simply selecting stars from these fields with a high number of observations ($n\geq30$ in this case) we can characterize the random error as a function of magnitude and filter.

\begin{figure*}
\begin{center}
 \includegraphics[width=18cm]{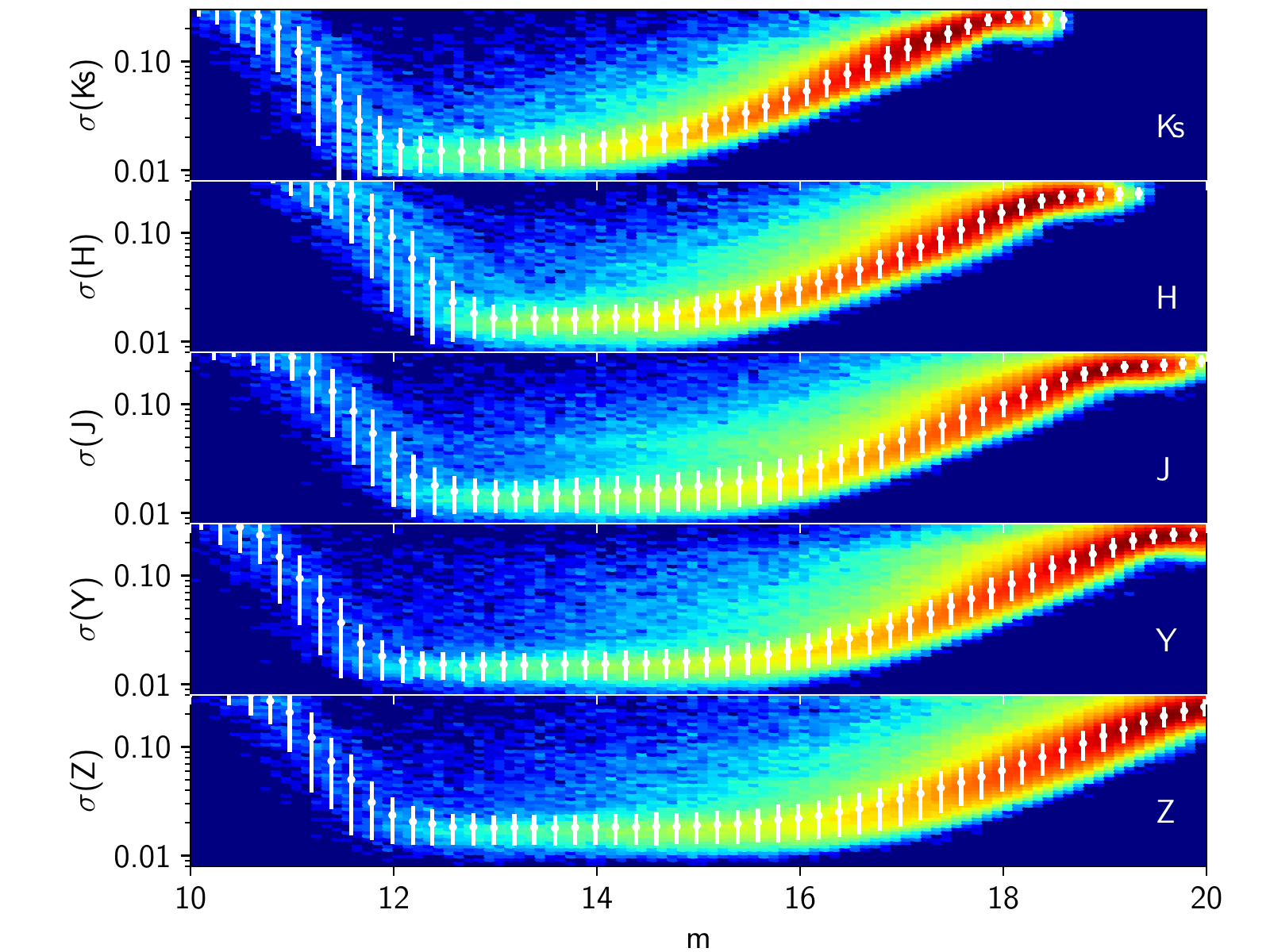}
 \caption{Magnitude standard deviation for repeated observations of stars in VISTA reference fields. Each of these stars have at least 30 measurements. White points represent a moving median, while the error bars mark $\pm1$ standard deviation. }
\label{stderr}
\end{center}
\end{figure*}

The results are summarised in Fig. \ref{stderr}. The error in magnitude, as parametrized by the standard deviation of repeated measurements of the same star, stays below 0.02 for about 5 magnitudes in the blue and 4 for the redder filters. Saturation typically occurs between 11.5 and 12.0 mags, and the total usable dynamic range, where the error is below $10\%$, is well over 7 mags for Z, Y and J, and 6 for H and K\,\!s. These estimations include all possible sources of error (atmospheric variation, radial distortion, etc.), but the reader should bear in mind that they are derived using stacked pawprints; the process of tiling these stacks introduces further systematics in the photometry that will be discussed later (Sect. \ref{res2d}). Reference fields are normally observed with a 5s integration time, and this also should be taken into account when using these results. 

\subsubsection{Putting VISTA in a Vega-like system}
\label{vistatovega}
\begin{figure}
\begin{center}
 \includegraphics[width=9cm]{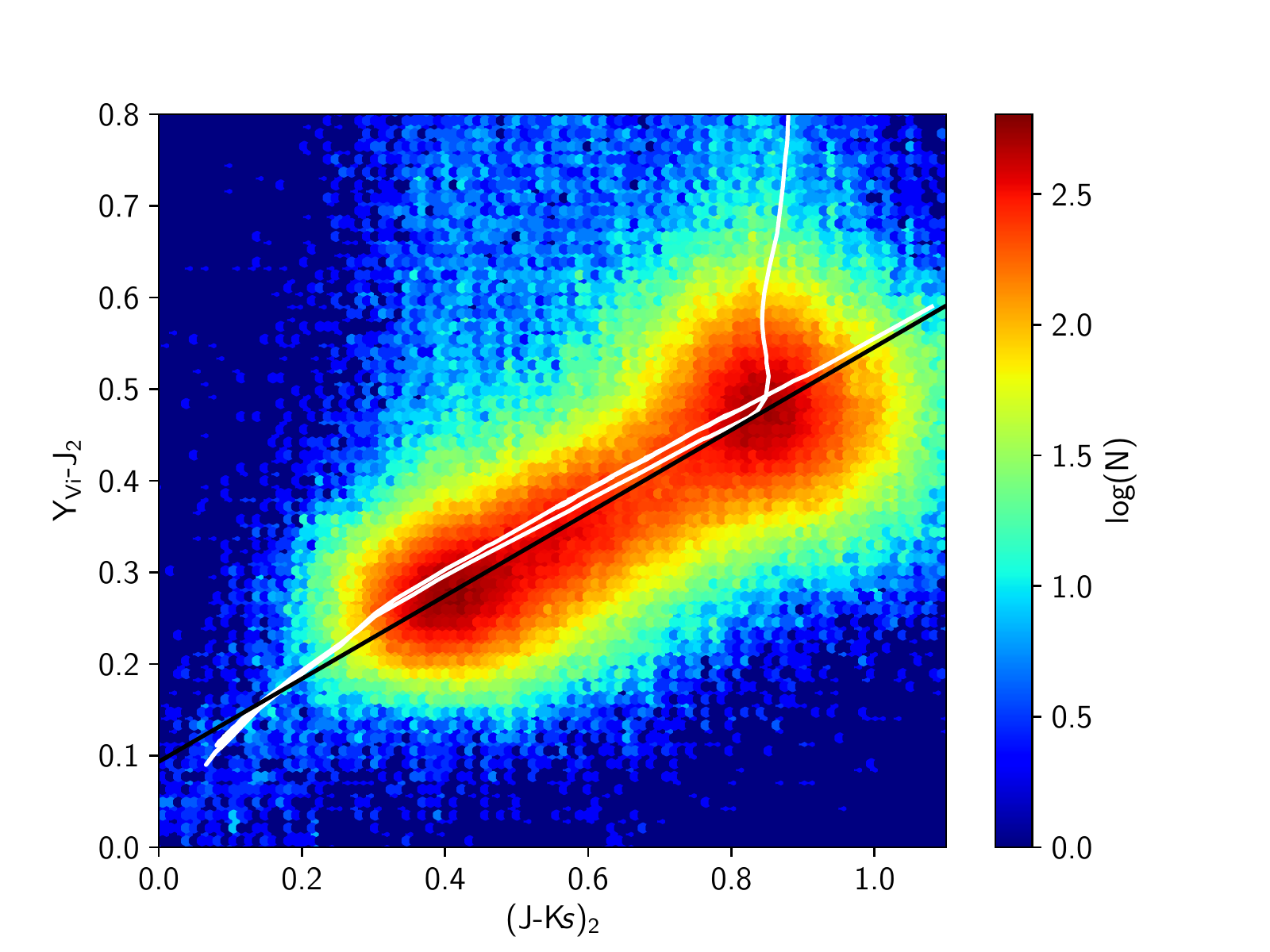}
 \caption{Calibration of the 2MASS-VISTA colour term in the Y band. The black line is fit from Eq. \ref{eq:yj}. The white line represents a 1 Gyr, Z=0.0 Padova isochrone.}
\label{yjiso}
\end{center}
\end{figure}

VISTA magnitudes are on a Vega-like system, and by definition, this requires that an unreddened A0V star has zero colour in all passbands. Zeropoints are derived from 2MASS magnitudes, also in a Vega system, so in theory calibrating against it should extend this property to VISTA. But the fact that the two system responses (detectors, filters, optics and atmosphere at the site) are differ introduces non-linear colour behaviour that could change this. As can be seen in Fig. \ref{yjiso}, for most of the range the $(Y_\mathrm{V}-J_\mathrm{2})$ colour is linear with $(J-K\mathrm{\!s})_\mathrm{2}$, but for blue stars (including an A0 population) there is a dip in $(Y_\mathrm{V}-J_\mathrm{2})$. This implies that magnitudes derived using the zeropoint calibration from Sec. \ref{2mequations}, that assume a completely linear colour to colour relation, will be slightly offset for A0V stars.

\begin{figure*}
\begin{center}
 \includegraphics[width=18cm]{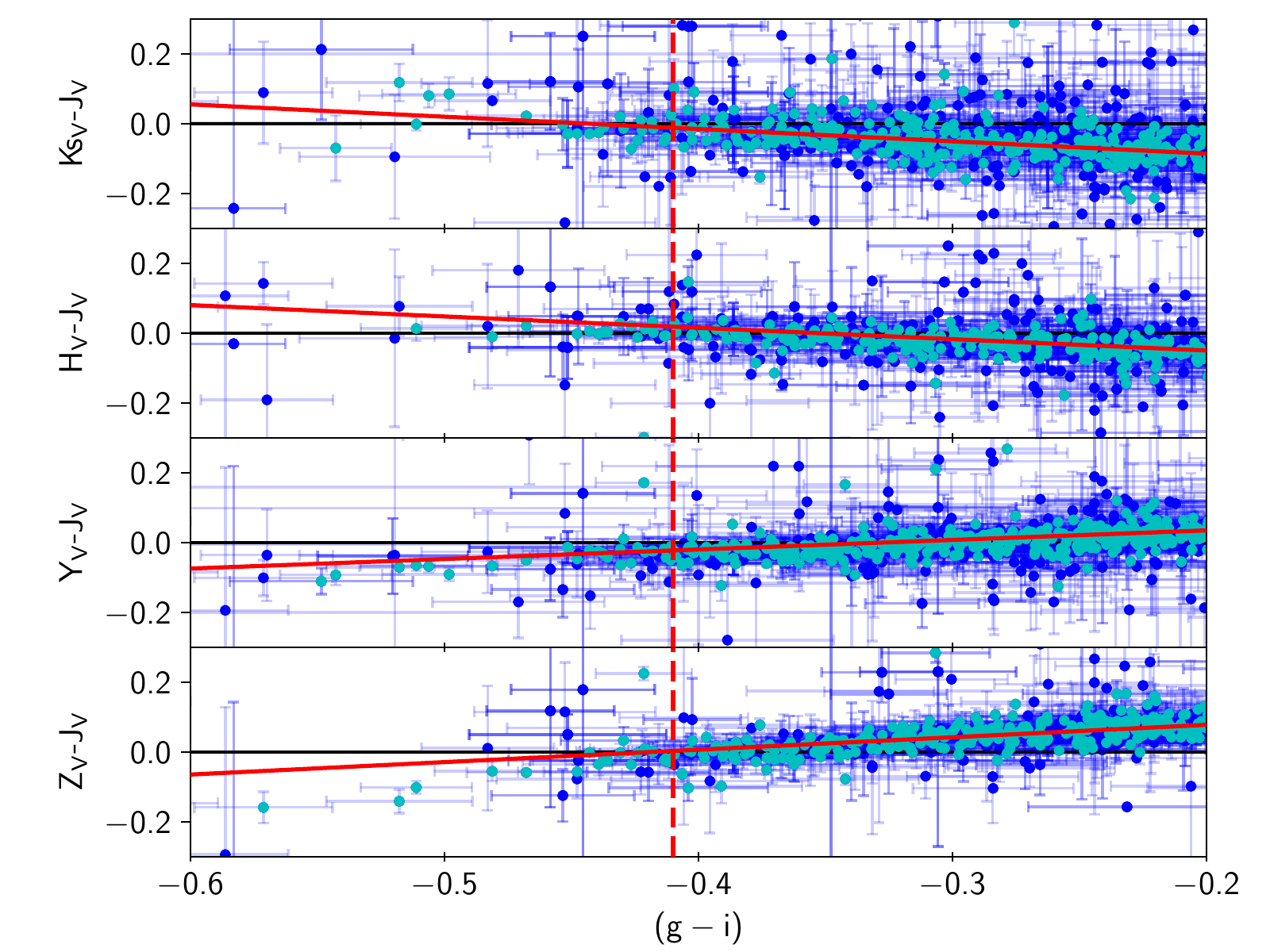}
 \caption{Determination of the zeropoint offset with respect to the J filter. In blue points we plot all the available A0 stars, while cyan marks those with good photometry used in the fit. The dashed line marks the intrinsic colour of an A0 star, according to \citet{fuk11}; the intercept between this and the linear fit, marked by the solid red line, determines the offset.}
\label{yzpa0}
\end{center}
\end{figure*}

To correct for possible offsets, we looked at the colours of A0V stars measured with VISTA. We selected Vega-like stars using the spectral classifications from SDSS DR9 \citep{ahn12}.
We can cross-match this set of stars and all the observations taken with VISTA\footnote{Using Q3C indexing \citep{kop06}}. In order to select only stars with good photometry, we adopt the selection criteria from sect. \ref{calto2mass}, except the $E(B-V)<0.1$ limit, since the apparent NIR colours of an A0 star should be linear with extinction, at least for the regions of relatively low extinction outside the Galactic plane we are using. The number of remaining A0 stars with measurements in all of  VISTA bands is of about a thousand, while about a hundred of these also have good 2MASS photometry (as this is our reference system). In Fig. \ref{yzpa0} we show the resulting colour-colour plots, using $(g-i)$ for SDSS and the $J$ band as reference for VISTA. It has the best spatial coverage (along with $K\mathrm{\!s}$), and at the same time is more useful to compare with 2MASS, as will be discussed later on. 

An example of these colour-colour diagrams can be seen in Fig. \ref{yzpa0}. The selected stars  occupy a colour range of about 0.6 in SDSS colours. To minimize the effect of reddening, we perform a robust linear fit between the respective NIR difference and $(g-i)$, and assume the offset for a Vega-like star is the value the fit yields for $(g-i)=-0.41$, the intrinsic colour of an A0V\footnote{SDSS magnitudes are in an AB system.} star according to \citet{fuk11}.

This procedure yields the colours for an A0V in the VISTA photometric system:
\begin{align}
\text{{\small VISTA col}}&\text{{\small ours }}\text{{\small of an A0V star}}\nonumber\\
Z_\mathrm{V}-J_\mathrm{V}&=0.004\pm0.005\label{zvjv}\\
Y_\mathrm{V}-J_\mathrm{V}&=-0.022\pm0.003\label{yvjv}\\
H_\mathrm{V}-J_\mathrm{V}&=0.019\pm0.003\label{hvjv}\\
K\mathrm{\!{s}_V}-J_\mathrm{V}&=-0.011\pm0.004\label{kvjv}
\end{align}

The colour offsets are quite modest, so native VISTA magnitudes are close, within $2\%$, of true Vega magnitudes. Now, comparing with 2MASS, that  has a good absolute calibration \citep{coh03}:

\begin{align}
\text{{\small VISTA-2MASS}}&\text{{\small \ colours }}\text{{\small of an A0V star}}\nonumber\\
(J_\mathrm{V}-J_\mathrm{2})&=0.005\pm0.015\label{jvj2}\\
(H_\mathrm{V}-H_\mathrm{2})&=0.029\pm0.014\\
(K\mathrm{\!{s}_V}-K\mathrm{\!{s}_2})&=0.04\pm0.02\label{kvk2}
\end{align}

As can be seen, there's no significant offset between both systems in $J$, and so it is reasonable to use this band to anchor our colour transformations. Despite the fact the initial SDSS sample is the same, the set that has good photometry in 2MASSS is only about one tenth of that for VISTA, and so the uncertainties in eqs. \ref{jvj2} to \ref{kvk2} are larger than those in eqs. \ref{zvjv} to \ref{kvjv}. We applied the same method to WFCAM photometry, as a sanity check:

\begin{align}
\text{{\small WFCAM-2MASS}}&\text{{\small \ colours }}\text{{\small of an A0V star}}\nonumber\\
(Z_\mathrm{W}-J_\mathrm{2})&=-0.02\pm0.09\label{eq:wz2j}\\
(Y_\mathrm{W}-J_\mathrm{2})&=0.04\pm0.03\label{eq:wy2j}\\
(J_\mathrm{W}-J_\mathrm{2})&=0.002\pm0.02\\
(H_\mathrm{W}-H_\mathrm{2})&=0.01\pm0.03\\
(K\mathrm{_W}-K\mathrm{\!{s}_2})&=0.05\pm0.03\label{eq:wk2k}
\end{align}

These results confirm the conclusion that WFCAM is already been on a Vega system. Being so, eqs. \ref{zvjv} to \ref{kvjv}, eqs. \ref{eq:wy2j} to \ref{eq:wk2k} and eqs. \ref{eq:wzeq} to \ref{eq:wkeq} should be compatible. This is true, within errors, for all bands.

\begin{figure}
\begin{center}
 \includegraphics[width=9cm]{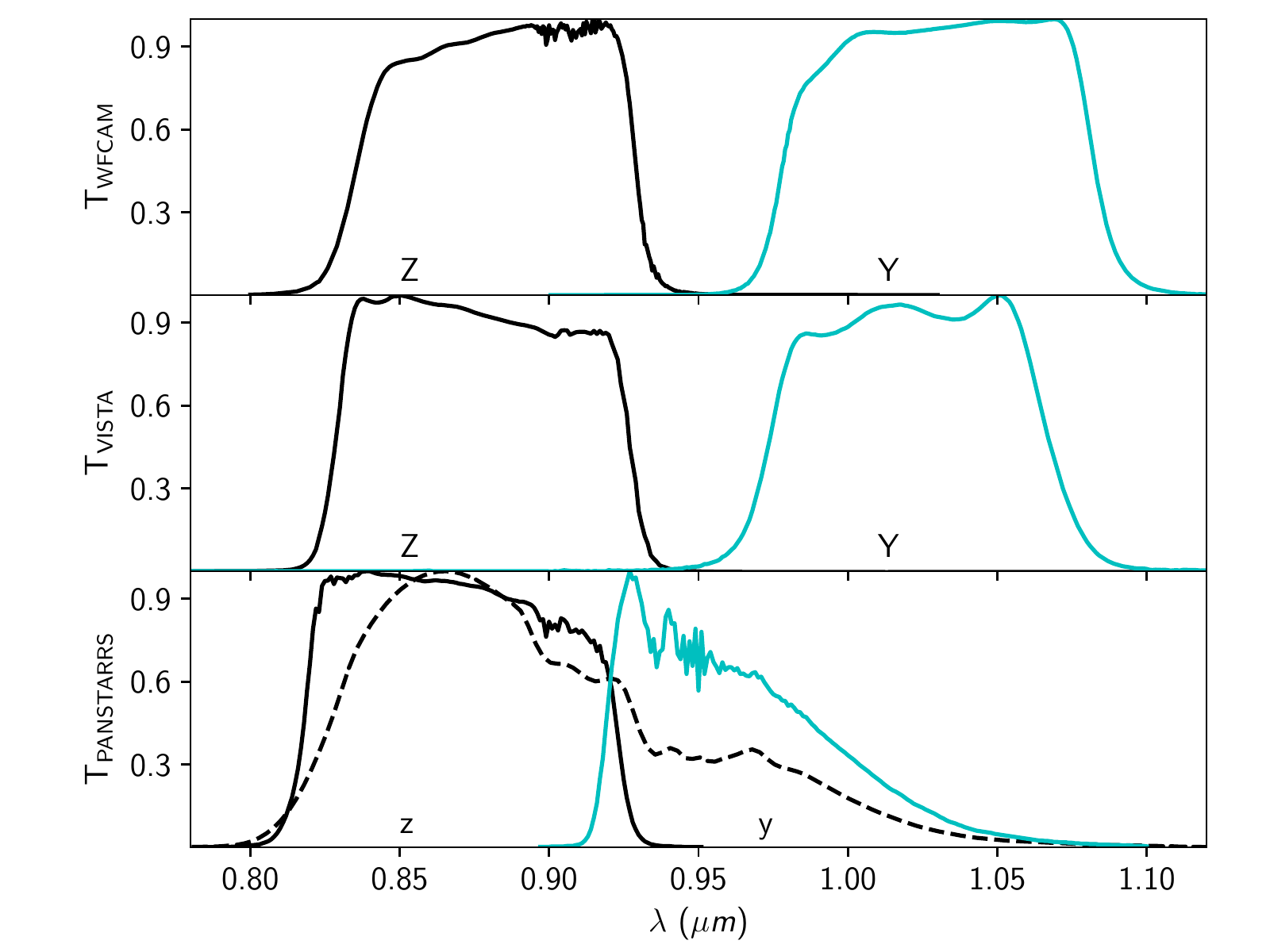}
 \caption{Transmission curves for VISTA and Pan-STARRS in the common wavelength range. The dashed line in the bottom panel is the $z$ filter from SDSS.}
\label{panfilt}
\end{center}
\end{figure}

The absolute calibration for the $Z$ and $Y$ bands was tested using the first data release from Pan-STARRS \citep{fle16}. While the $z$ filter from SDSS does not allow for a straightforward comparison with VISTA, Pan-STARRS is much more suitable for this (see Fig. \ref{panfilt}).

To compare VISTA with Pan-STARRS, that uses an AB system, we first put VISTA into Vega magnitudes with eqs. \ref{zvjv} to \ref{kvjv} and then into a AB magnitudes using the offsets from Appendix \ref{abmag}. Once done, we repeated the same calculation as in Fig. \ref{yzpa0} obtaining
\begin{align}
Z_\mathrm{V}-z_\mathrm{P}&=0.034\pm0.012\\
Y_\mathrm{V}-y_\mathrm{P}&=0.003\pm0.009
\end{align}

These result offer good agreement between both systems, particularly taking into account that even if similar, both filter sets differ (in particular $y$) and that the Vega to AB transformations from Appendix \ref{abmag} are also folded into these differences.

From these results we can draw a few conclusions about VISTA photometry as offered to the community:
\begin{enumerate}
\item The colour term for an A0V star between 2MASS and VISTA found here are in concordance with the absolute 2MASS calibration from \citet{coh03}, with significant offsets in H and K$\mathrm{\!s}$.
\item As WFCAM is already set in a Vega system using 2MASS, our method yields no colour terms when compared with 2MASS, not even for the H band, as the $3\%$ has already been accounted for (H09).
\item To transform VISTA magnitudes into a true Vega system, only small offsets are needed (about $2\%$ in Y and H, Eqs. \ref{yvjv} and \ref{hvjv}).
\end{enumerate}

It is important to highlight that, by default, {\bf these offsets are not applied to any CASU derived data products}. It is the responsability of the user, if she wishes, to apply eqs. \ref{zvjv} to \ref{kvjv} in order to put VISTA in a true Vega system. For this, as the absolute offset in $J_\mathrm{V}$ is negligible (eq. \ref{jvj2}), the coefficients from eqs. \ref{zvjv} to \ref{kvjv} should be subtracted from VISTA magnitudes.

\subsection{The effects of tiling}

The process of grouting is necessary when generating photometric catalogues over tiled images. It compensates the variations in seeing between stacked pawprints, and removes the spatial systematics that these introduce in the photometry. This can be checked by comparing the magnitudes measured in a tiled image, both pre and post grouting, with the pristine photometry from the corresponding stacks. As each one of these stacks has a short exposure time and comes with its own aperture correction, they are not affected by variations in seeing.
\begin{figure}
\begin{center}
 \includegraphics[width=9cm]{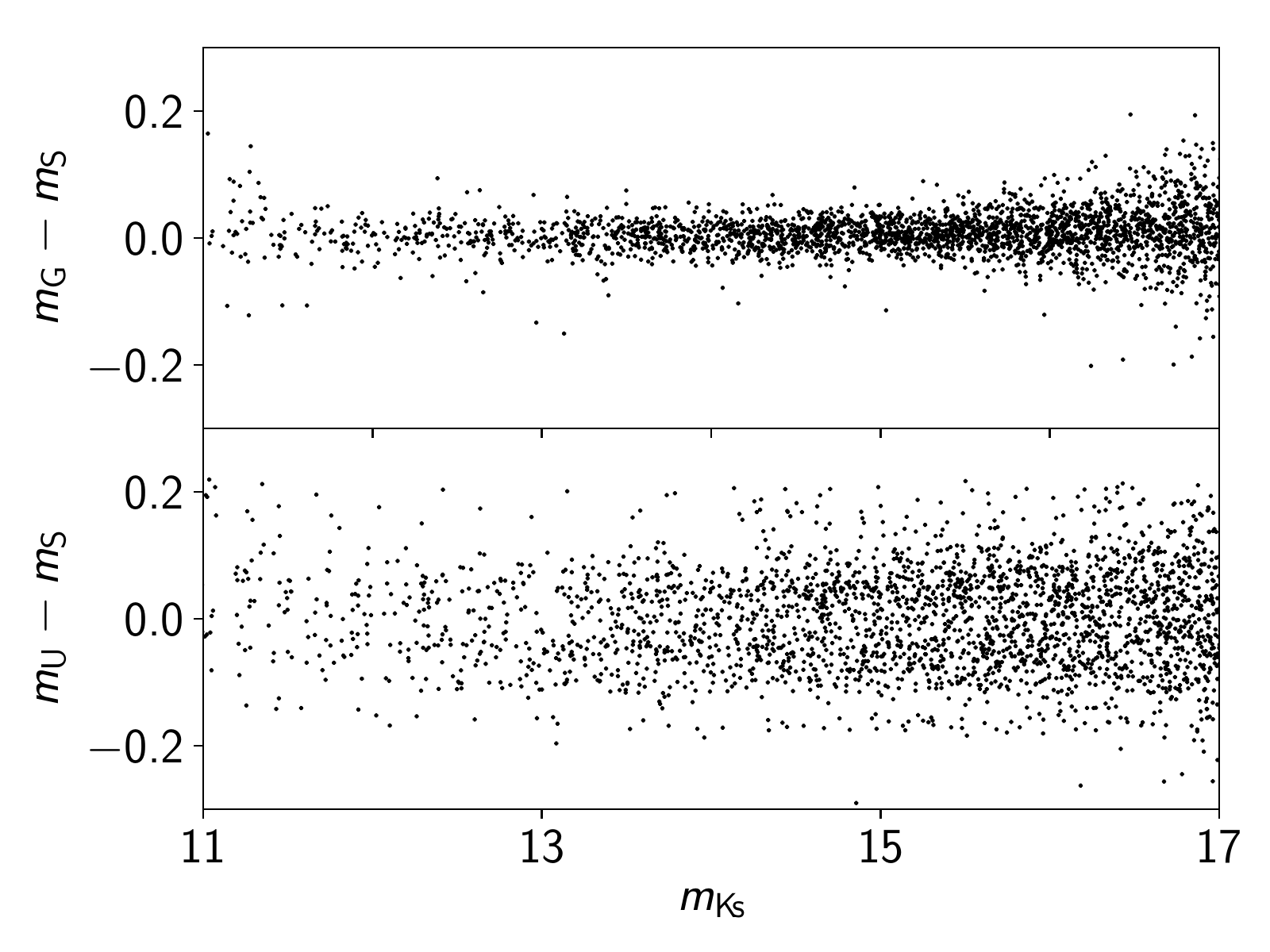}
 \caption{Difference in magnitude between an ungrouted ($m_\mathrm{U}$) and grouted ($m_\mathrm{G}$) tile and the corresponding stacks. As can be seen in the top panel, while grouting corrects most of the systematics, there is some dispersion introduced by the stack to tile conversion.}
\label{grout2}
\end{center}
\end{figure}

The result of this can be seen in Fig. \ref{grout2}. While the grouted magnitudes correlate much better with the stacked catalogues, there is still some dispersion, of about $2\%$. This increase in flux error is intrinsic to the mosaicing process, and users that want the most precise photometry are advised to use stack images and catalogues.

\subsection{Spatial distribution of the residuals}
\label{res2d}
\begin{figure}
\begin{center}
 \includegraphics[width=9cm]{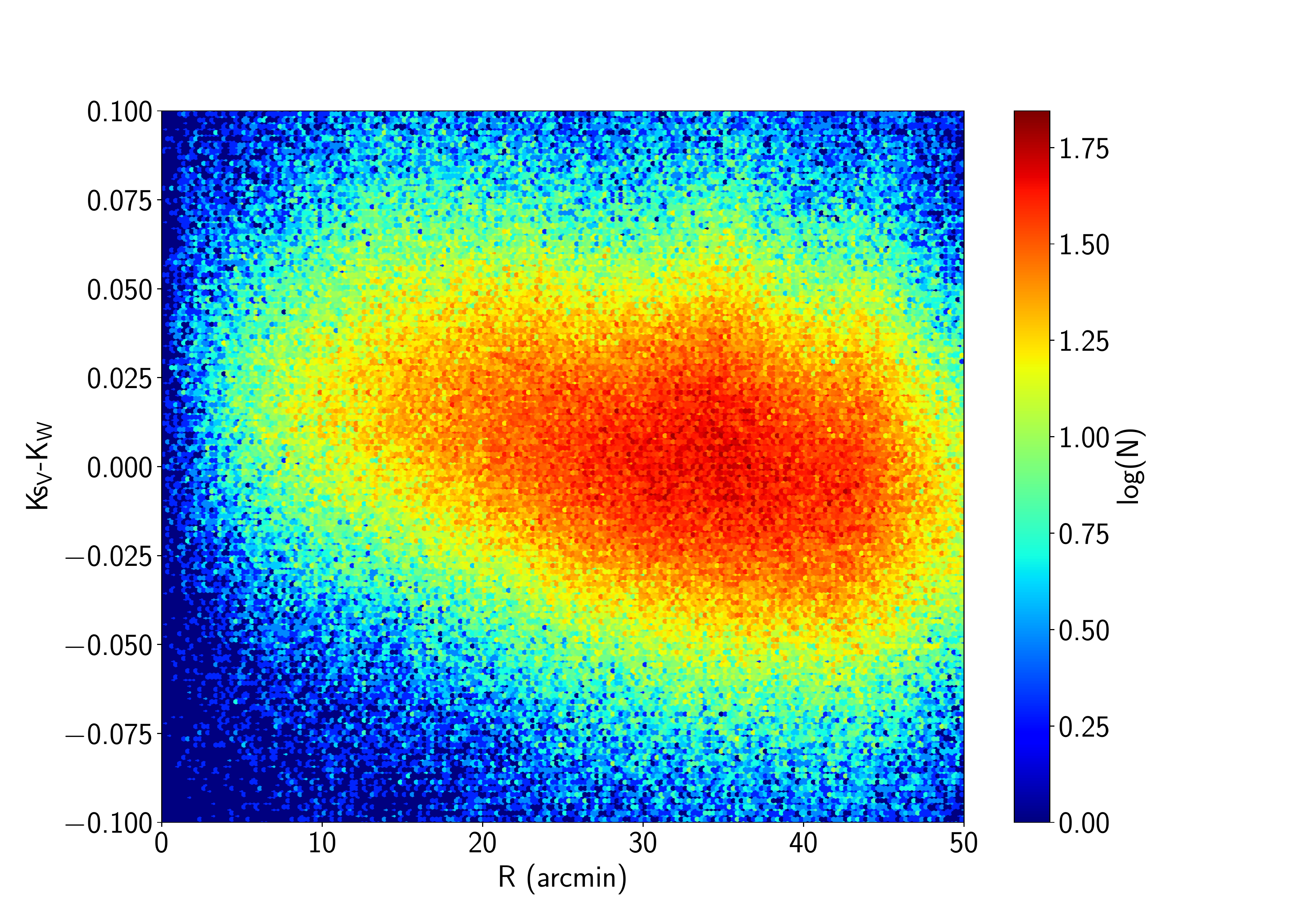}
 \caption{Variation of the magnitude difference between VISTA and WFCAM as a function of radius from the centre of the tile.}
\label{2dall_rad}
\end{center}
\end{figure}

\begin{figure}
\begin{center}
 \includegraphics[width=9cm]{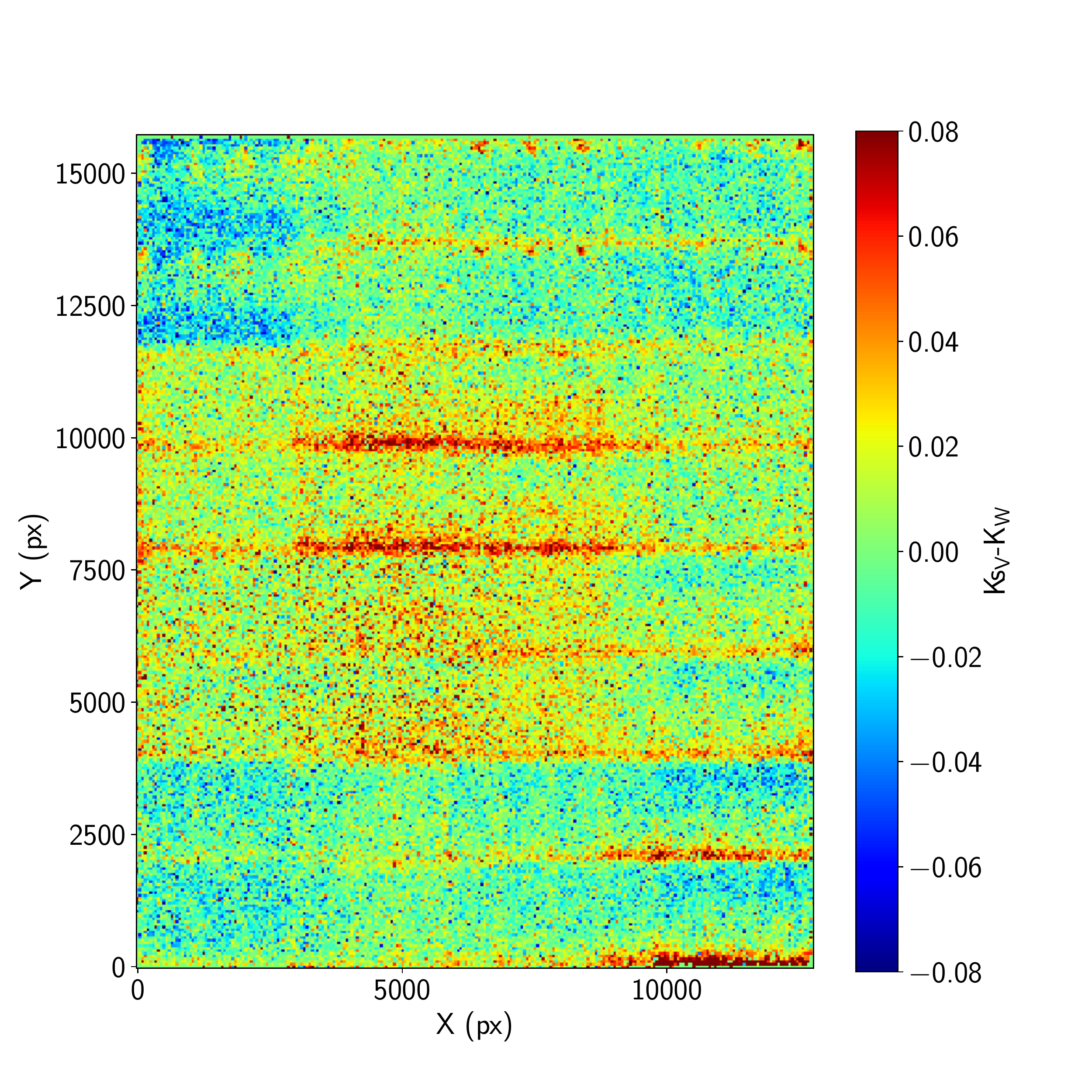}
 \caption{Median magnitude offset in the $K$ band between VISTA and WFCAM. Each pixel in the image corresponds to 50$\times$50 pixels in the detector array plane.}
\label{2dall}
\end{center}
\end{figure}

The sky coverage of the detector array in VIRCAM is not complete \citep{sut15}, so to generate a continuous image of the sky a $2\times3$ pointing pattern is used. Although the order of the pointings may change between observations, the offsets between them are always the same. This implies that we can define a coordinate system that all the tiles from VISTA share. Once this common reference frame is established, we can see if the photometric errors derived in sect. \ref{phores} have a spatial dependence. In Fig. \ref{2dall_rad} we plot the variation of $m_\mathrm{VISTA}-m_\mathrm{WFCAM}$ (for all the available tiles) with radius, as measured from the centre of the tile. As can be seen, although the rms of the error is typically around $3\%$, the magnitude difference has some spatial structure. This can be explored further by reconstructing the (x,y) distribution. For simplicity we will do this in virtual tile pixel space and calculating the median of $m_\mathrm{VISTA}-m_\mathrm{WFCAM}$ in regions of 50$\times$50 pixels. We plot the result of this procedure in Fig. \ref{2dall}.

It is clear from the plot that, while the majority of the tile area sits around $\Delta m\sim0.0$, there is a pattern to where this is not the case. In particular, several narrow horizontal bands with $\Delta m\sim0.05$, and a large cross-like pattern with $\Delta m\sim0.03$ stand up. The bands seem to align themselves with the detector boundaries; we can check if this is the case by repeating the same analysis but using instead of tiles, stacked pawprints, where the detectors retain their identity.

\begin{figure}
\begin{center}
 \includegraphics[width=9cm]{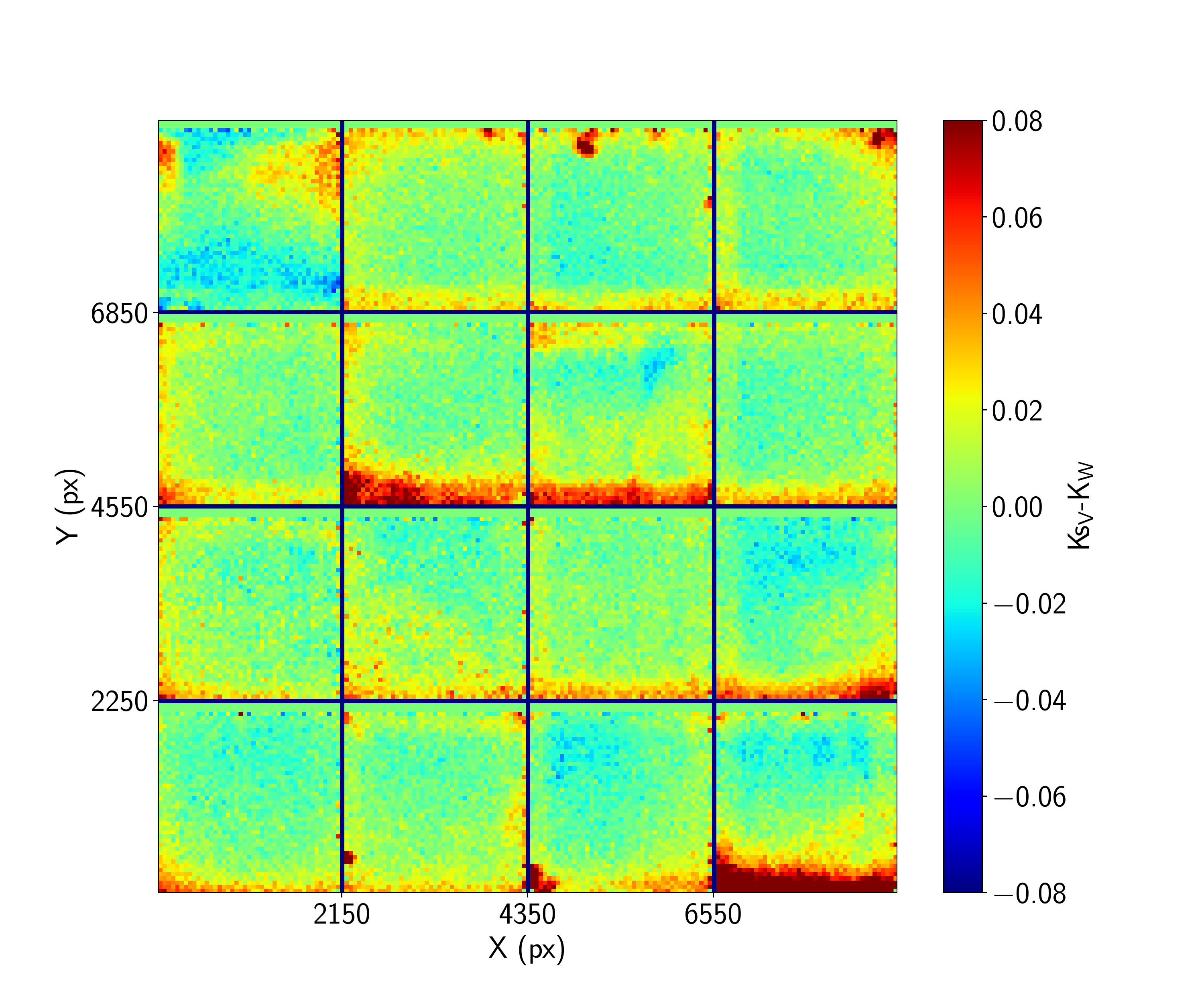}
 \caption{Same as Fig. \ref{2dall}, but using stacks instead of tiles. For clarity, the insterspacing between detectors has been reduced to 50 pixels. Individual detectors are wider than $2048\times2048$ as stacks are built from two or more individual exposures taken with a small offset.}
\label{2dstack}
\end{center}
\end{figure}

This can be seen in Fig. \ref{2dstack}. Approximately, the bottom tenth of every detector shows a magnitude offset of a few percent. The origin of this is not clear. It seems to have a weak dependency on wavelength, with the amplitude of the offset decreasing towards the blue, but this might also be related to exposure time, as the average time the detectors are exposed also changes with filter. Also, the readout electronics for the detectors in VIRCAM is concentrated in the bottom part of each array, coincident with the observed offset area. This could point to a temperature effect, but further investigations are needed.  

\begin{figure}
\begin{center}
 \includegraphics[width=9cm]{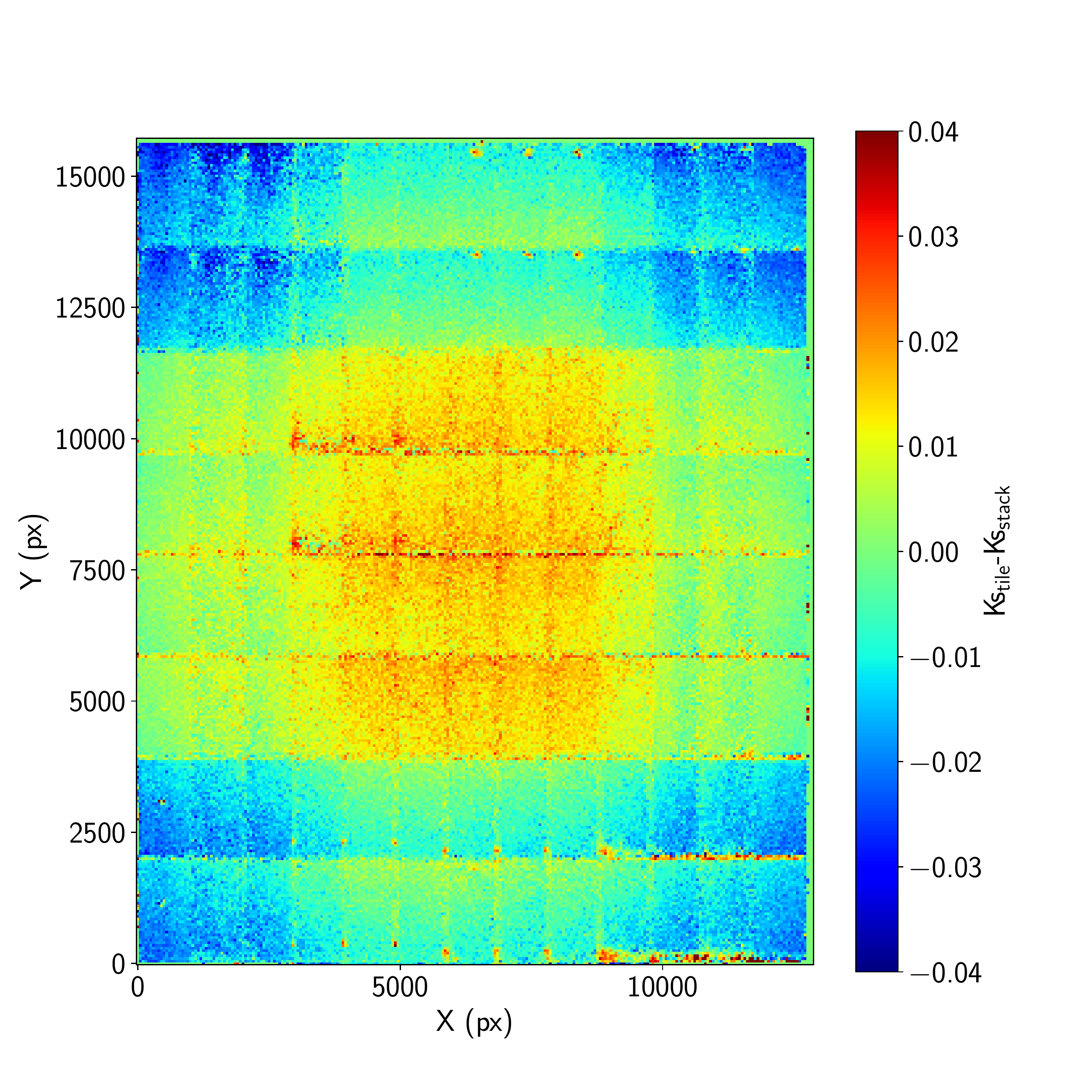}
 \caption{Difference between photometry taken from stacked pawprints and tiles}
\label{tilestack}
\end{center}
\end{figure}

As has been outlined in sect. \ref{insmag}, when transforming from stacks into tiles, the projection takes care of correcting the radial distortion present in the array plane. If this distortion is not properly accounted and several stacks are combined into a tile, a cross-like pattern like the one observed in Fig. \ref{2dall} is expected. In fact, if we compare the photometry measured from tiled images from that of the equivalent stacked frames, we obtain the distribution plotted in Fig. \ref{tilestack}. The fact that this effect is still present in tiled images (and not in stacks, as there is no obvious radial trend in Fig. \ref{2dstack}) points to a problem in the tiling procedure at the $3\%$ level that needs to be corrected in future version of the pipeline.

These corrections are an analytical function of the pointing pattern that goes into a tile, and therefore they can be retroactively applied to photometry catalogues with relative ease. In the case of tiled images, although the fix itself is simple too, applying it retroactively would be too expensive in computer time. Therefore, the proposed solution is the release of a tool that would allow users to correct tiled images if they need to.

\section{Conclusions}

\begin{enumerate}

\item The CASU pipeline photometric calibration of VIRCAM data into the VISTA photometric system is based on 2MASS photometry for stars observed in each VIRCAM pointing.

\item In regions of moderate Galactic Extinction ($E(B-V)<$5.0), we apply an $E(B-V)$ dependent correction to the photometry, which results in a VISTA photometric calibration that is independent of the reddening. For the bluest filters, regions under greater extinction may show residual uncalibrated colour terms. In $JHK\!\mathrm{s}$ this correction is stable until $E(B-V)\sim7$

\item This technique has achieved a photometric precision, for stacked pawprints, better than $2\%$ in the $YJHK\!\mathrm{s}$ filters and $3\%$ in $Z$ for most of the dynamic range of the detector, as measured by repeat measurements of stars in the VISTA reference fields.

\item Using these data we are able to investigate the long term sensitivity of VISTA, and monitor the decline in performance of the VISTA mirrors (with a Silver coating), and improvement in the stability of the mirrors (with an Aluminium coating), and the jumps in sensitivity when the VIRCAM entrance window is cleaned.

\item Tiling handles a number of important potential sources of error in the VISTA photometry: variable PSFs between the detectors, and variations in the seeing between the pawprints that make up a tile. We show that the average differences between pawprint and tile catalogue photometry are zero, but there is a spatial dependency of the difference, that can reach $4\%$ for the pixels farther away from the centre of the detector array (Fig. \ref{tilestack}). This is due to the incorrect application of the distortion correction, and will be taken care in future data releases (see Appendix \ref{datavers}). 

\item We derive the internal offsets that put VISTA on a Vega system, and the colour transformations and offsets between VISTA and 2MASS, and VISTA and WFCAM.

\end{enumerate}

\section*{Acknowledgments}

Our thanks go to the VISTA consortium for building, and the staff of ESO for operating, VISTA and VIRCAM.

PCH acknowledges support from the STFC via a Consolidated Grant to the Institute of Astronomy, Cambridge.

This publication makes use of data products from the Two Micron All Sky Survey, which is a joint project of the University of Massachusetts and the Infrared Processing and Analysis Center/California Institute of Technology, funded by the National Aeronautics and Space Administration and the National Science Foundation.

Funding for the SDSS and SDSS-II has been provided by the Alfred P. Sloan Foundation, the Participating Institutions, the National Science Foundation, the U.S. Department of Energy, the National Aeronautics and Space Administration, the Japanese Monbukagakusho, the Max Planck Society, and the Higher Education Funding Council for England. The SDSS Web Site is http://www.sdss.org/.

The SDSS is managed by the Astrophysical Research Consortium for the Participating Institutions. The Participating Institutions are the American Museum of Natural History, Astrophysical Institute Potsdam, University of Basel, University of Cambridge, Case Western Reserve University, University of Chicago, Drexel University, Fermilab, the Institute for Advanced Study, the Japan Participation Group, Johns Hopkins University, the Joint Institute for Nuclear Astrophysics, the Kavli Institute for Particle Astrophysics and Cosmology, the Korean Scientist Group, the Chinese Academy of Sciences (LAMOST), Los Alamos National Laboratory, the Max-Planck-Institute for Astronomy (MPIA), the Max-Planck-Institute for Astrophysics (MPA), New Mexico State University, Ohio State University, University of Pittsburgh, University of Portsmouth, Princeton University, the United States Naval Observatory, and the University of Washington.

The Pan-STARRS1 Surveys (PS1) and the PS1 public science archive have been made possible through contributions by the Institute for Astronomy, the University of Hawaii, the Pan-STARRS Project Office, the Max-Planck Society and its participating institutes, the Max Planck Institute for Astronomy, Heidelberg and the Max Planck Institute for Extraterrestrial Physics, Garching, The Johns Hopkins University, Durham University, the University of Edinburgh, the Queen's University Belfast, the Harvard-Smithsonian Center for Astrophysics, the Las Cumbres Observatory Global Telescope Network Incorporated, the National Central University of Taiwan, the Space Telescope Science Institute, the National Aeronautics and Space Administration under Grant No. NNX08AR22G issued through the Planetary Science Division of the NASA Science Mission Directorate, the National Science Foundation Grant No. AST-1238877, the University of Maryland, Eotvos Lorand University (ELTE), the Los Alamos National Laboratory, and the Gordon and Betty Moore Foundation.

This work makes use of matplotlib \citep{hun07}, numpy and scipy.








\appendix

\section{Fitting the 2MASS colour term}
\label{ap:colour}
\subsection{Which colour to use?}
\begin{figure}
\begin{center}
\includegraphics[width=8.5cm]{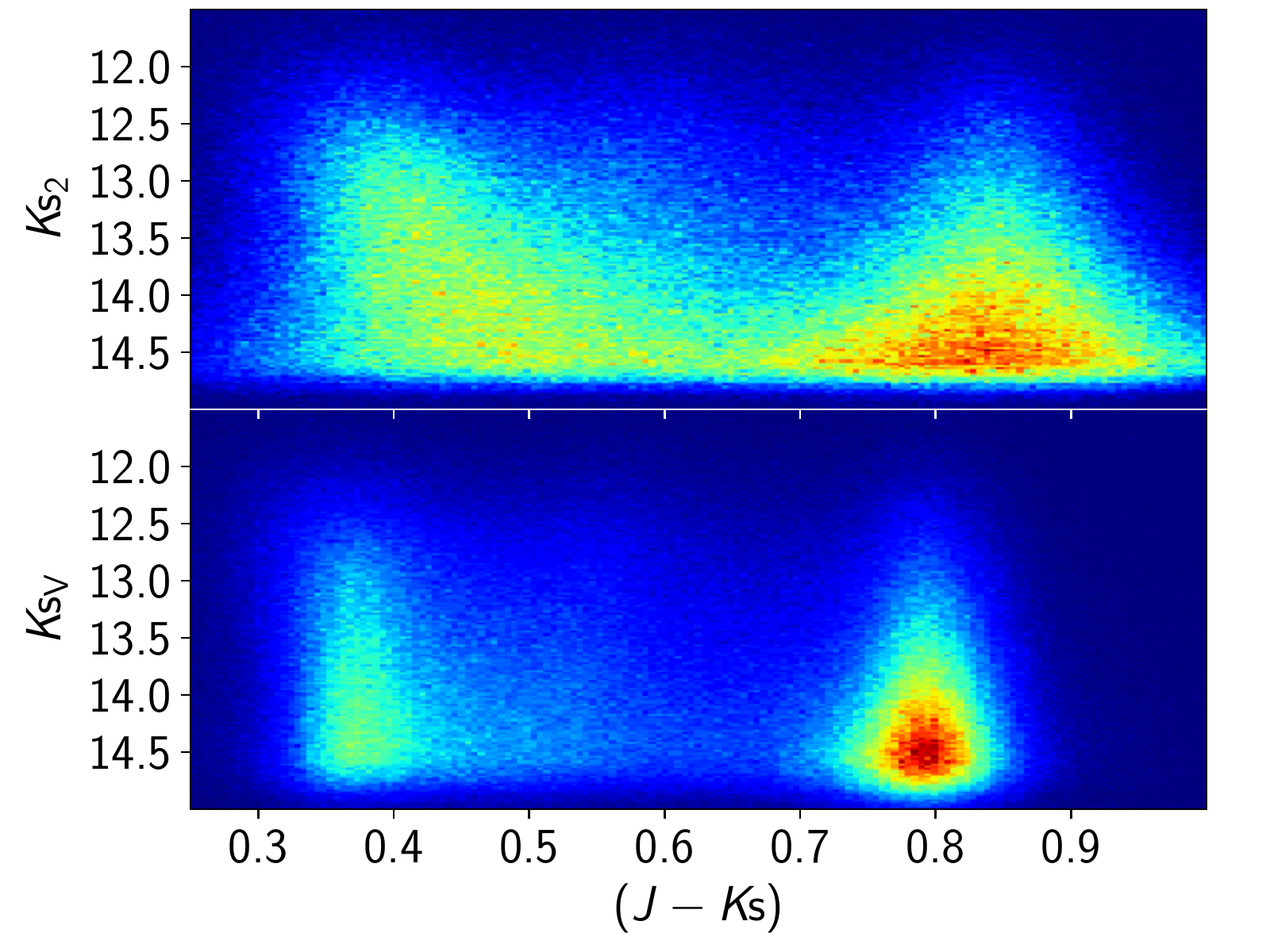}
\caption{Colour-magnitude diagram in both the VISTA (bottom) and 2MASS (top) systems for the sources common with 2MASS used in Sect. \ref{2mequations}.}
\label{fig:cmd_vista}
\end{center}
\end{figure}

In theory, the 2MASS to VISTA calibration can be parametrized as a function of $(J-H)_\mathrm{2}$ or $(J-K\!\mathrm{s})_\mathrm{2}$. In practice, there are nuances that tip the balance towards $(J-K\!\mathrm{s})_\mathrm{2}$. In particular, sources with good photometry in both surveys and in high Galactic latitude fields, mainly come from two populations (see Fig. \ref{fig:cmd_vista}), late red dwarfs (at $(J-K\!s)\sim0.8$) being the most populous of them.

\begin{figure}
\begin{center}
\includegraphics[width=8.5cm]{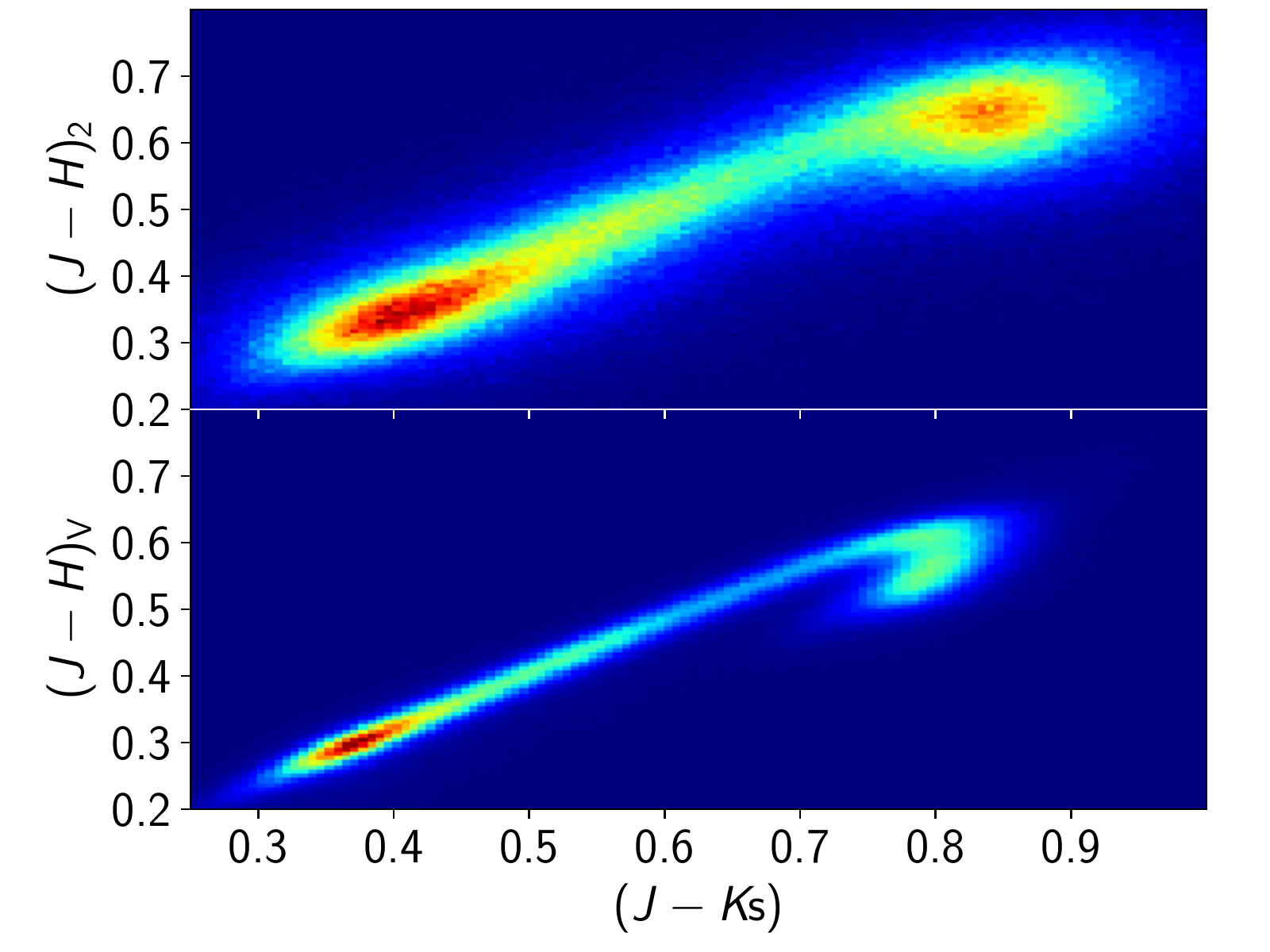}
\caption{Colour-colour diagram in both the VISTA (bottom) and 2MASS (top) systems for the sources common with 2MASS used in Sect. \ref{2mequations}.}
\label{fig:ccd_vista}
\end{center}
\end{figure}

The near infrared SED of very late red dwarfs is marked by the presence of water absorption, that affect the $H$, $Y$ and $Z$ bands severely. The $J$ band is less influenced by them and $K\!\mathrm{s}$ is barely sensitive to them. This implies that the locus of late dwarfs on a colour-colour diagram shows a clear "hook" (see Figs. \ref{fig:ccd_vista} and \ref{yzpa0}), as these stars shift to the blue in $(J-H)$ (and to the red in $(Y_\mathrm{V}-J_\mathrm{2})$). While  the small errors in VISTA show this effect clearly, in 2MASS the result is that the $(J-H)$ distribution is blurred and skewed to the blue, making it much more difficult to model. On top of this, the populations visible in Fig. \ref{fig:cmd_vista} tend to be more spread out in $(J-K\!s)$ than in $(J-H)$ allowing for easier fits.

Based on these, we opt to calibrate the VISTA system using $(J-K\!s)_2$.

\subsection{Fitting for the colour term}
\label{fitcolterm}
We obtain the 2MASS colour terms by fitting a linear equation of the form
\begin{equation}
J_\mathrm{V}-J_\mathrm{2}=A+B\cdot(J-K\mathrm{\!s})_\mathrm{2}
\end{equation}
for all the VISTA filters, although as we use instrumental magnitudes, the $A$ term is discarded. We do these fits on a monthly basis. We split a colour-colour plot like the one in fig. \ref{yzpa0} in intervals of 0.03 mag in $(J-K\!s)_2$ and obtain the median and a robust estimate of the dispersion within. We fit the medians weighted with the dispersion and number of stars in each interval. To restrict ourselves to the range where the relation between colours is linear, we only fit the interval $0.35\leq(J-K\!s)_2\leq0.85$ in all bands except $Z$, where a more conservative cut of $(J-K\!s)_2\leq0.8$ is applied due to the stronger presence of water bands for red stars.

\section{VISTA extinction coefficients}
\label{cofext}
In order to transform from the $E(B-V)$ colour excesses provided by \citet{sche98} into the expected extinction at a given band i, $A_\mathrm{i}$, we need the relative to absolute extinction coefficient $R_\mathrm{i}$, as 
\begin{equation}
R_\mathrm{i}=\frac{A_\mathrm{i}}{E(B-V)}
\end{equation}

These coefficients are notoriously complicated to calculate, so for VISTA we use as a starting point those provided by \citet{scha11}. Assuming $R_\mathrm{V}=3.1$, we interpolate the coefficients provided in Table 6 of the aforementioned work to the effective wavelength of the VISTA and 2MASS filters. This calculation results in the values
\begin{align*}
&(R_\mathrm{Z},R_\mathrm{Y},R_\mathrm{J},R_\mathrm{H},R_\mathrm{K\!s})_\mathrm{V}=(1.395,1.017,0.705,0.441,0.308)\\
&(R_\mathrm{J},R_\mathrm{H},R_\mathrm{K\!s})_2=(0.725, 0.448, 0.307)
\end{align*}
With these values we can estimate the $D_\mathrm{i}$ coefficients needed in eq. \ref{eqn3}: 
\begin{equation*}
(D_\mathrm{Z},D_\mathrm{Y},D_\mathrm{J},D_\mathrm{H},D_\mathrm{K\!s})=(0.349,0.131,-0.008,-0.013,0.003)
\end{equation*}
and generate a set of zeropoints for fields at a wide range of galactic latitudes (and so reddening) observed under good atmospheric conditions. The throughput of the VISTA system should not depend on the average $E(B-V)$ value of the field, so the extinction coefficients are refined until any remaining dependencies are corrected. The values that show the best behaviour are:

\begin{align}
&D_\mathrm{Z} = 0.264\\
&D_\mathrm{Y} = 0.103\\
&D_\mathrm{J} = 0.000\\
&D_\mathrm{H} = 0.000\\
&D_\mathrm{K\!s} = 0.005
\end{align}

It should be noted that these values account mostly for the differences in the effects of extinction when comparing VISTA and 2MASS. As the differences between the two sets of ($J,H,K\mathrm{\!s}$) filters are small, so are these coefficients.

\section{VISTA data versioning}
\label{datavers}

CASU keeps an internal versioning system that follows major changes in the data processing pipeline, following the form {\bf 1.X.Y}, where successive {\bf X} values mark a major change in the processing and successive {\bf Y} instead are reserved for minor updates like bugs and changes in deep sky masks for those surveys that use them.

VISTA data are made available under two major releases, {\bf 1.3} and {\bf 1.5}. 

\subsection{Version 1.3}
The pipeline used for this version was verified using only science verification data and early science data. Therefore, all the issues raised in Sect. \ref{overper} apply to this version. As the available calibration data was more limited, the colour equations used to derive the zeropoints are also different from those in Sect. \ref{2mequations}; for version {\bf 1.3} the conversion between 2MASS and VISTA follows:
\begin{align}
&Z_{\rm V} = J_{\rm 2} + 1.025\cdot(J-H)_{\rm 2}\\
&Y_{\rm V} = J_{\rm 2} + 0.610\cdot(J-H)_{\rm 2}\\
&J_{\rm V} = J_{\rm 2} - 0.077\cdot(J-H)_{\rm 2}\\
&J_{\rm V} = J_{\rm 2} - 0.065\cdot(J-K\!{\rm s})_{\rm 2}\\
&H_{\rm V} = H_{\rm 2} + 0.032\cdot(J-H)_{\rm 2}\\
&K\!{\rm s_V} = K\!{\rm s_2} + 0.010\cdot(J-K\!{\rm s})_{\rm 2}
\end{align}

Those equations derived for $(J-K\!{\rm s})_{\rm 2}$ offer a reasonable agreement with the new values from Sect. \ref{2mequations}.

\subsection{Version 1.5}
The biggest change introduced for version {\bf 1.5} is the use of eqs. \ref{eq:zj} to \ref{eq:ks} to determine the zeropoints. As they depend on $(J-K\!{\rm s})_{\rm 2}$ for all filters, the transformations are more stable with respect to Galactic latitude than those from version {\bf 1.3}, as the slope of the colour term is less affected by the relative abundance of late dwarfs in the sample used.

The offsets to Vega determined in sect. \ref{vistatovega} were measured with data from version {\bf 1.3}, but the changes in absolute calibration are expected to be minimal. We can check nonetheless, re-calculating the zeropoints for all the data used in sec. \ref{vistatovega}. The differences are:
\begin{align}
&Z_{\rm 1.3} - Z_{\rm 1.5}= -0.03\pm0.01\\
&Y_{\rm 1.3} - Y_{\rm 1.5}= 0.018\pm0.004\\
&J_{\rm 1.3} - J_{\rm 1.5}= -0.0200\pm0.0008\\
&H_{\rm 1.3} - H_{\rm 1.5}= 0.0067\pm0.0003\\
&K\!{\rm s_{1.3}} - K\!{\rm s_{1.5}}=0.0106\pm0.0007
\end{align}

Apart from changing the colour equations, the grouting procedure has also been improved for version {\bf 1.5}. The main result from this change is the correction of the 2D pattern seen in Fig. \ref{2dstack}.

\section{Conversion of flux into magnitudes}
\label{abmag}

The processing philosophy is to preserve the image and catalogue data as counts, and to document all the required calibration information in the file headers. Thus recalibration of the data requires only changes to the headers, and these headers can be reingested into the WSA
without the need to reingest the full tables. For readers accessing the flat files (catalogues and images) rather than the WSA database products, we document the methods for converting the fluxes into magnitudes and calibrating the photometry.

\begin{equation}
m=ZP-2.5log_{\rm 10}(\frac{f}{t})-A-k(\chi-1)
\label{eqn:photapp}
\end{equation}

where $ZP$ is the zeropoint for the frame (keyword: MAGZPT in the FITS header), $f$ is the flux within the chosen aperture (e.g. column: APER\_FLUX\_3), $t$ is the exposure time for each combined integration (keyword: EXP\_TIME), and $A$ is the appropriate aperture correction (e.g. keyword: APCOR3). The final term deals with the extinction correction, where $k$ is the extinction coefficient (EXTINCTION) and is equal to 0.05 magnitudes/airmass in all filters,
and $\chi$ is the airmass (keywords: AMSTART, AMEND).

While stacks retain the original ZPN projection, tile images and catalogues have been transformed to a TAN coordinate system. This implies that the radial correction outlined on Sect. \ref{insmag} has already been applied to the pixel fluxes, and therefore the value for $f$ is the one straight from the catalogue. In the case of stacks, $f$ in eq. \ref{eqn:photapp} must be substituted by $f_{\rm cor}$, as specified in  eq. \ref{eq:distortcor}.

It is possible to transform from the VISTA system to AB magnitudes. Firstly, we need to move from the VISTA internal magnitudes to true Vega magnitude applying eqs. \ref{zvjv} to \ref{kvjv}, and then following \citet{hew06}, apply these offsets:
\begin{align}
&Z_\mathrm{AB}-Z_\mathrm{V}=0.502\\
&Y_\mathrm{AB}-Y_\mathrm{V}=0.600\\
&J_\mathrm{AB}-J_\mathrm{V}=0.916\\
&H_\mathrm{AB}-H_\mathrm{V}=1.366\\
&K\!s_\mathrm{AB}-K\!s_\mathrm{V}=1.827
\end{align}

\section{Relevant source and image parameters}
\label{headers}

\begin{table*}
\caption{Table of source parameters generated by the VIRCAM pipeline and
  written to the FITS catalogue
  products, and an accompanying short description for each.}
\begin{tabular}{|p{1cm}|p{3.0cm}|p{11cm}|}
\hline
1 & Sequence number & Running number for ease of reference, in strict
order of image detections.\\
2 & Isophotal flux & Standard definition of summed flux within detection isophote, apart from detection filter is used to define pixel connectivity and hence which pixels to include. This helps to reduce edge effects for all isophotally derived parameters.\\
3 & X coord & Intensity-weighted isophotal centre-of-gravity in X.\\
4 & Error in X & Estimate of centroid error.\\
5 & Y coord & Intensity-weighted isophotal centre-of-gravity in Y.\\
6 & Error in Y & Estimate of centroid error.\\
7 & Gaussian sigma & Derived from the three intensity-weighted second
moments. The equivalence to a generalised elliptical Gaussian
distribution is used to derive:

Gaussian sigma = $(\sigma_a^2+\sigma_b^2)^{1/2}$\\

8 & Ellipticity & ellipticity = $1.0-\sigma_a/\sigma_b$\\
9 & Position angle & position angle = angle of ellipse major axis wrt x axis\\
10--16& Areal profile 1 

Areal profile 2

Areal profile 3

Areal profile 4

Areal profile 5

Areal profile 6

Areal profile 7
& Number of pixels above a series of threshold
levels relative to local sky. Levels are set at T, 2T, 4T, 8T ... 128T
where T is the threshold. These can be thought of as a poor
man's radial profile. For deblended, i.e. overlapping
images, only the first areal profile is computed and the rest are set
to -1.\\
17 & Areal profile 8 & For blended images this parameter is used to flag the start of the sequence of the deblended components by setting the first in the sequence to 0\\
18 & Peak height & In counts relative to local value of sky - also zeroth order aperture flux\\
19 & Error in peak height & \\
20--45 & Aperture flux 1 

Error in flux 

Aperture flux 2

Error in flux 

Aperture flux 3

Error in flux 

Aperture flux 4

Error in flux 

Aperture flux 5

Error in flux 

Aperture flux 6

Error in flux 

Aperture flux 7

Error in flux 

Aperture flux 8

Error in flux 

Aperture flux 9

Error in flux 

Aperture flux 10

Error in flux 

Aperture flux 11

Error in flux 

Aperture flux 12

Error in flux 

Aperture flux 13

Error in flux 
& These are a series of different radius
soft-edged apertures designed to adequately sample the curve-of-growth
of the majority of images and to provide fixed-sized aperture fluxes
for all images. The scale size for these apertures is selected by
defining a scale radius $\sim<$FWHM$>$ for site+instrument. In the case of
VIRCAM this "core" radius (rcore) has been fixed at 1.0 arcsec for
convenience in inter-comparison with other datasets. A 1.0 arcsec
radius is equivalent to 2.5 pixels for non-interleaved data, 5.0
pixels for 2x2 interleaved data, and 7.5 pixels for 3x3 interleaved
data. In $\sim$1 arcsec seeing an rcore-radius aperture contains roughly
2/3 of the total flux of stellar images. (The rcore
parameter is user specifiable and hence is recorded in the output
catalogue FITS header.)

The aperture fluxes are sky-corrected integrals (summations) with a
soft-edge (i.e. pro-rata flux division for boundary pixels). However,
for overlapping images the fluxes are derived via simultaneously
fitted top-hat functions, to minimise the effects of crowding. Images
external to the blend are also flagged and not included in the large
radius summations.

Aperture flux 3 is recommended if a single number is required to
represent the flux for ALL images - this aperture has a radius of
rcore.

Starting with parameter 20 the radii are: (1) $1/2 \times$rcore, (2)
$1/\sqrt2 \times$rcore, (3) rcore, (4) $\sqrt2 \times$rcore, (5) $2
\times$rcore, (6) $2\sqrt2 \times$rcore, (7) $4 \times$rcore, (8)
$5 \times$rcore, (9) $6 \times$rcore, (10) $7 \times$rcore,
(11) $8 \times$rcore, (12) $10 \times$rcore, (13) $12 \times$rcore.

Note $4 \times$rcore contains $\sim$99\% of PSF flux.

The apertures beyond Aperture 7 are for generalised galaxy photometry.

Note larger apertures are all corrected for pixels from overlapping
neighbouring images.

The largest aperture has a radius 12$\times$rcore ie. $\sim$24 arcsec diameter.

The aperture fluxes can be combined with later-derived aperture
corrections for general purpose photometry and together with parameter
18 (the peak flux) give a simple curve-of-growth measurement which
forms the basis of the morphological classification scheme.\\ 
46 & Petrosian radius & $r_p$ as defined in Yasuda et al. 2001 AJ 112
1104 \\
47 & Kron radius & $r_k$ as defined in Bertin and Arnouts 1996 A\&A Supp
117 393 \\
48 & Hall radius & $r_h$ image scale radius eg. Hall \& Mackay 1984 MNRAS
210 979\\\hline
\end{tabular}
\end{table*}

\begin{table*}
\begin{tabular}{|p{1cm}|p{3.0cm}|p{11cm}|}
\hline
49 & Petrosian flux & Flux within circular aperture to $k \times r_p$; $k$ = 2\\
50 & Error in flux & \\
51 & Kron flux & Flux within circular aperture to $k \times r_k$; $k$ = 2\\
52 & Error in flux& \\
53 & Hall flux & Flux within circular aperture to $k \times r_h$; $k$ = 5;
alternative total flux\\ 
54 & Error in flux& \\
55 & Error bit flag & Bit pattern listing various processing error flags\\
56 & Sky level & Local interpolated sky level from background tracker\\
57 & Sky $rms$ & local estimate of $rms$ in sky level around image\\
58 & Child/parent & Flag for parent or part of deblended deconstruct (redundant since only deblended images are kept)\\
59-60 & RA 

DEC
& RA and Dec explicitly put in columns for overlay programs
that cannot, in general, understand astrometric solution coefficients
- note r*4 storage precision accurate only to $\sim50$mas. Astrometry can
be derived more precisely from WCS in header and XY in parameters 5 \&
6\\
61 & Classification & Flag indicating most probable morphological classification: eg. -1 stellar, +1 non-stellar, 0 noise, -2 borderline stellar, -9 saturated\\
62 & Statistic & An equivalent N(0,1) measure of how stellar-like an
image is, used in deriving parameter 61 in a "necessary but not
sufficient" sense. Derived mainly from the curve-of-growth of flux
using the well-defined stellar locus as a function of magnitude as a
benchmark (see Irwin et al. 1994 SPIE 5493 411 for more details).\\ \hline
\end{tabular}
\end{table*}

\begin{table*}
\caption{Relevant photometric parameters measured by the pipeline and
  written to the FITS headers. These values are computed per-detector
  and stored in the headers for each image and catalogue
  extension. The names by which the parameters are stored in the VIRCAM
Science Archive tables are also given.}
\begin{tabular}{|p{2.5cm}|p{2.5cm}|p{9cm}|}
\hline
FITS keyword & WSA parameter & Description\\\hline
AMSTART & amStart & Airmass at start of observation\\
AMEND & amEnd & Airmass at end of observation\\
PIXLSIZE & pixelScale & $[$arcsec$]$ Pixel size\\
SKYLEVEL & skyLevel& $[$counts/pixel$]$ Median sky brightness\\
SKYNOISE & skyNoise & $[$counts$]$ Pixel noise at sky level\\
THRESHOL & thresholdIsoph & $[$counts$]$ Isophotal analysis threshold\\
RCORE & coreRadius & $[$pixels$]$ Core radius for default profile fit\\
SEEING & seeing & $[$pixels$]$Average stellar source FWHM\\
ELLIPTIC & avStellarEll & Average stellar ellipticity (1-b/a)\\
APCORPK & aperCorPeak & $[$magnitudes$]$ Stellar aperture correction -- peak height\\
APCOR1 & aperCor1 & $[$magnitudes$]$ Stellar aperture correction -- core/2 flux\\
APCOR2 & aperCor2 & $[$magnitudes$]$ Stellar aperture correction -- core/$\sqrt2$ flux \\
APCOR3 & aperCor3 & $[$magnitudes$]$ Stellar aperture correction -- core flux \\
APCOR4 & aperCor4 & $[$magnitudes$]$ Stellar aperture correction --
$\sqrt2 \times$ core flux \\
APCOR5 & aperCor5 & $[$magnitudes$]$ Stellar aperture correction -- $2
\times$ core flux\\
APCOR6 & aperCor6 & $[$magnitudes$]$ Stellar aperture correction --
$2\sqrt2 \times$
core flux\\
APCOR7 & aperCor7 & $[$magnitudes$]$ Stellar aperture correction -- $4
\times$ core flux \\
MAGZPT & photZPExt &  $[$magnitudes$]$ Photometric ZP for default extinction\\
MAGZRR & photZPErrExt & $[$magnitudes$]$ Photometric ZP error\\
EXTINCT & extinctionExt &$[$magnitudes$]$ Extinction coefficient\\
NUMZPT & numZPCat & Number of 2MASS standards used\\
NIGHTZPT & nightZPCat & $[$magnitudes$]$ Average photometric ZP for the filter for
the night\\
NIGHTZRR & nightZPErrCat & $[$magnitudes$]$ Photometric ZP $\sigma$ for the filter for the night\\
NIGHTNUM & nightZPNum & Number of ZPs measured for the filter for the night\\
\hline
\end{tabular}
\end{table*}


\bsp	
\label{lastpage}
\end{document}